\documentclass[aps,preprint,superscriptaddress,nofootinbib,floatfix,showpacs]{revtex4-1}
\pdfoutput=1

\usepackage{epsfig}
\usepackage{dcolumn}
\usepackage{bm}
\usepackage{amssymb}
\usepackage{amsmath}
\usepackage{hyperref}
\usepackage{xcolor}

\newcommand{\ord}[1]{\mathcal{O} \left( #1 \right)}

\begin{document}
\title{Baseline predictions of elliptic flow and fluctuations at the RHIC Beam Energy Scan using response coefficients}

\author{S. Rao}
\affiliation{Rutgers University, Piscataway, NJ USA 08854}
\author{M. Sievert}
\affiliation{Rutgers University, Piscataway, NJ USA 08854}
\affiliation{University of Illinois at Urbana-Champaign, Urbana, IL 61801, United States}
\author{J. Noronha-Hostler}
\affiliation{Rutgers University, Piscataway, NJ USA 08854}
\affiliation{University of Illinois at Urbana-Champaign, Urbana, IL 61801, United States}

\date{\today}
\begin{abstract}
Currently the RHIC Beam Energy Scan is exploring a new region of the Quantum Chromodynamic phase diagram at large baryon densities that approaches nuclear astrophysics regimes.  This provides an opportunity to study relativistic hydrodynamics in a regime where the net conserved charges of baryon number, strangeness, and electric charge play a role, which will significantly change the theoretical approach to simulating the baryon-dense Quark-Gluon Plasma. Here we detail many of the important changes needed to adapt both initial conditions and the medium to baryon-rich matter.  Then, we make baseline predictions for the elliptical flow and fluctuations based on extrapolating the physics at LHC and top RHIC energies to support future analyses of where and how the new baryon-dense physics causes these extrapolations to break down. First we compare eccentricities across beam energies, exploring their underlying assumptions; we find the the extrapolated initial state is predicted to be nearly identical to that at AuAu $\sqrt{s_{NN}}=200$ GeV.  Then the final flow harmonic predictions are based on linear+cubic response.  We discuss preliminary STAR results in order to determine the implications that they have for linear+cubic response coefficients at the lowest beam energy of AuAu $\sqrt{s_{NN}}=7$ GeV. 

\end{abstract}

\maketitle

%
\section{Introduction}
%

Since the Quark-Gluon Plasma was first measured experimentally in the early 2000's, the field has progressed significantly in understanding this nearly perfect fluid.  Quantitative theory-to-experimental-data comparisons are now possible at the Large Hadron Collider (LHC) and Relativistic Heavy Ion Collider (RHIC) using event-by-event viscous relativistic hydrodynamics  \cite{Gardim:2012yp,Gardim:2012im,Heinz:2013bua,Shen:2015qta,Niemi:2015qia,Niemi:2015voa,Noronha-Hostler:2015uye,Gardim:2016nrr,Gardim:2017ruc,Eskola:2017bup,Giacalone:2017dud,Gale:2012rq,Bernhard:2016tnd,Giacalone:2016afq,Zhu:2016puf,Zhao:2017yhj,Zhao:2017rgg,Alba:2017hhe,Noronha-Hostler:2019ytn,sievert:2019zjr,Bozek:2011if,Bozek:2012gr,Bozek:2013ska,Bozek:2013uha,Kozlov:2014fqa,Zhou:2015iba,Mantysaari:2017cni,Weller:2017tsr}.  At top beam energies where the number of baryons to anti-baryons is approximately equal (i.e. the baryon chemical potential is $\mu_B \approx 0$), a small shear viscosity to entropy density ratio of $\eta/s\approx 0.05-0.2$ has been extracted from these comparisons.  It has been shown that the Quantum Chromodynamics (QCD) equation of state obtained from lattice simulations can describe the data well \cite{Pratt:2015zsa,Moreland:2015dvc,Alba:2017hhe,Auvinen:2018uej}.  However, some of the largest remaining uncertainties are the nature of the initial state immediately after two heavy ions collide and how this state evolves to approach hydrodynamic behavior. Recent years have seen success in constraining the initial state by comparisons to elliptical flow distributions \cite{Renk:2014jja} and multiparticle cumulants \cite{Giacalone:2017uqx}. 

Once the ultra-relativistic kinematics of top RHIC and LHC energies are relaxed to lower beam energies, this well-established paradigm of heavy-ion collisions can be significantly modified.  Notably, the assumption of an equal number of baryons and anti-baryons at mid-rapidity is violated, leading to nonzero baryon chemical potential ($\mu_B> 0$) at lower beam energies.  These changes can significantly affect all the stages of the heavy-ion collision, and the theoretical modeling must be adapted accordingly.  A brief list of some of the most obvious changes are:
\begin{itemize}
\item \textbf{Initial conditions: } 
	At top collider energies, the mid-rapidity region is characterized by very small $x \sim p_T / \sqrt{s}$ which is dominated by gluons, leading to an initial state with $\mu_B \approx 0$.  Lower beam energies correspond to larger values of $x$, leading to a non-negligible contribution of baryon stopping ($\mu_B > 0$) and new valence quark degrees of freedom in the initial state \cite{Itakura:2003jp}.  Additionally, the violation of high-energy ``eikonal'' kinematics introduces a number of challenging corrections which are usually power-suppressed, including a finite overlap time of the colliding nuclei \cite{Shen:2017bsr}.  Generally, it is thought that hydrodynamics must start at later time for lower beam energies \cite{Karpenko:2015xea}.
\item \textbf{Equation of State: } 
	At finite baryon densities not only must all three conserved charges be considered (baryon number B, strangeness S, and electric charge Q) \cite{Borsanyi:2018grb,Noronha-Hostler:2019ayj,Monnai:2019hkn}, but also the cross-over phase transition curves downward to lower temperatures \cite{Bazavov:2018mes,Critelli:2017oub,Borsanyi:2018grb} and there may possibly be a critical point, which needs to be incorporated into the equation of state \cite{Parotto:2018pwx}.
\item \textbf{Transport coefficients: } 
	At $\mu_B=0$ generally shear $\eta/s$ and bulk viscosity $\zeta/s$ are considered (although certain models also incorporate second order transport coefficients).  However at $\mu_B>0$, not only 
	%
	do these quantities
	vary with the chemical potentials $\mu_B , \mu_S , \mu_Q$  \cite{Demir:2008tr,Denicol:2013nua,Kadam:2014cua} (and possibly see effects from a critical point \cite{Monnai:2016kud}), but also new transport coefficients that describe the diffusion of the conserved charges (BSQ) must be considered.  \cite{Rougemont:2015ona,Rougemont:2017tlu,Greif:2017byw,Denicol:2018wdp,Martinez:2019bsn}.  These new transport coefficients can also be affected by the presence of a critical point.
\item \textbf{Freeze-out and Chemical Equilibrium:} 
	Lower beam energies generally also appear to lead to lower chemical freeze-out temperatures \cite{Alba:2014eba,Borsanyi:2014ewa,Bellwied:2018tkc,Andronic:2017pug}.  
\item \textbf{Critical Fluctuations:} 
	If a critical point exists at low beam energies, critical fluctuations must also be included within hydrodynamical models \cite{Stephanov:1999zu,Jiang:2015hri,Mukherjee:2016kyu,Stephanov:2017ghc,Nahrgang:2018afz,An:2019osr}, although  a consensus of the exact description of this within relativistic viscous hydrodynamics has not yet been reached. 
\end{itemize}
While some hydrodynamic models are beginning to incorporate these effects \cite{Du:2019obx,Denicol:2018wdp,Batyuk:2017sku}, this is still very much a work in progress.  Therefore, it will likely take some years before it is possible to do the same systematic theory-to-experimental-data comparisons that have already been performed at top collider energies where $\mu_B \approx 0$.  Additionally, much of the experimental data at the lowest beam energies still have large enough error bars to warrant waiting for the completion of Beam Energy Scan II analysis.  

In the meantime, it can be quite instructive to perform baseline calculations of bread-and-butter observables such as flow harmonics in order to see how far our current knowledge of hydrodynamics at $\mu_B \approx 0$ can take us.  In this paper we focus specifically on elliptical flow, which is known to arise from a combination of linear and cubic response from the initial state \cite{Noronha-Hostler:2015dbi} and we use our best knowledge at high beam energies to extrapolate downwards.  We know that these extrapolations to low beam energies must eventually break down significantly, as the physics of the low-energy baryon-rich regime discussed above begins to play an important role, but exactly where and how these new effects set in is unclear.  Therefore, an observation of large systematic deviations from these baseline predictions can indicate the onset of these new physical mechanisms, and a quantitative analysis of such deviations from the baseline can help to disentangle whether the new physics arises from changes in the initial or final state. 

The rest of this paper is organized as follows.  In Sec.~\ref{sec:Framework} we lay out the theoretical framework we will use to explore the beam energy dependence of the standard picture of heavy-ion collisions.  We lay out our choice of models in Sec.~\ref{subsec:Model}, the estimator + residual approach to predicting the final flow harmonics in Sec.~\ref{subsec:estimators}, and the initial state comparison we will pursue in Sec.~\ref{sec:TrentoMCKLN}.  In Sec.~\ref{sec:initialstate} we explore the energy dependence associated with the initial state, both through its explicit dependence on the experimental cross section and through potential secondary dependence on changing model parameters.  In Sec.~\ref{sec:kappas} we extract the linear+cubic response coefficients and residuals using two different methods and determine their energy dependence.  In Sec.~\ref{sec:extract} we detail three possible choices for how to extrapolate from top RHIC and LHC energies down to lower beam energies.  In Sec.~\ref{sec:results} we use the established framework to extrapolate down the baseline predictions for $v_2 \{2\}$ and $v_2 \{4\} / v_2 \{2\}$ for AuAu collisions at $54~\mathrm{GeV}$, $27~\mathrm{GeV}$, and $7~\mathrm{GeV}$.  In particular, we explore in Sec.~\ref{subsec:extractdata} the possibility of extracting the response coefficients directly from data.  Finally, we conclude in Sec.~\ref{sec:concl} with a summary of our main results.

%
\section{Framework}
\label{sec:Framework}
%

%
\subsection{Model}
\label{subsec:Model}
%

The standard paradigm of heavy-ion collisions requires two stages of model input: one for generating the initial conditions at the time $\tau_0$ at which hydrodynamic evolution begins, and another for the hydrodynamics and freeze-out.  
Our working hydrodynamic model consists of the 2+1D event-by-event relativistic viscous hydrodynamics code, v-USPhydro \cite{Noronha-Hostler:2013gga, Noronha-Hostler:2014dqa}.  v-USPhydro utilizes the smoothed-particle implementation of hydrodynamics, and we use the parameters $\tau_0=0.6$ fm, $\eta/s\sim0.05$, $\zeta/s=0$, $T_{FO}=150$ MeV.  For the equation of state, we use the most up-to-date Lattice QCD extractions PDG16+/EOS21 from \cite{Alba:2017hhe}, and we treat freeze-out using the Cooper-Frye prescription.  This setup has been compared extensively to data across many beam energies and system sizes \cite{Alba:2017hhe, Giacalone:2017dud, sievert:2019zjr}.  

For the initial conditions we will primarily use the Trento model \cite{Moreland:2014oya,Bernhard:2016tnd} to set the initial energy density.  The ``standard" Trento parameter set $p=0$, $k=1.6$, and $\sigma=0.51$ has been shown to describe the experimental data well \cite{Moreland:2014oya,Bernhard:2016tnd,Alba:2017hhe,Giacalone:2017dud} and appears to produce similar event geometries to the IP-Glasma model \cite{Schenke:2012wb,Gale:2012rq} and EKRT \cite{Niemi:2015qia}.  We relax these three Trento parameters at lower beam energies to explore the flexibility of the model framework to describe low-energy nuclear collisions.  

In order to extract the response coefficients we compare Trento+vUSPhydro calculations from \cite{Alba:2017hhe,sievert:2019zjr,Katz:2019qwv} versus the MC-KLN+v-USPhydro calculations from \cite{Noronha-Hostler:2016eow,Betz:2016ayq,Katz:2019fkc} (where we use the MC-KLN code from \cite{Drescher:2007ax,Drescher:2006ca} to generate the initial conditions). One should note that the MC-KLN+v-USPhydro used an outdated equation of state and lower freeze-out temperature than is typical but is still interesting to compare to see how sensitive the response coefficients are to a completely different medium parameterizations.  

%
\subsection{Flow Estimators}
\label{subsec:estimators}
%

In hydrodynamics, multiparticle correlations arise from the independent emission of particles at freeze-out which are mutually correlated with the event geometry (event plane).  The anisotropies of the single-particle distribution $\frac{dN}{d^2 p}$ are characterized by the complex flow vectors
\begin{equation}
\bm{V_n} \equiv v_n e^{i n \psi_n}
\end{equation}
which can be expressed in terms of a magnitude $v_n$ and phase $\psi_n$ (event plane angle) for the $n^{th}$ order harmonic.  These single-particle flow vectors then form the building blocks for the measured multiparticle correlations in a hydrodynamic picture.  Similarly, the initial state immediately following a heavy ion collision can be characterized using a complex eccentricity vector
\begin{equation}
\bm{\mathcal{E}_n} \equiv \varepsilon_n e^{i n \phi_n}
\end{equation}
with magnitude $\varepsilon_n$ and phase $\phi_n$.  

The near-perfect fluidity of the quark-gluon plasma has been shown \cite{Teaney:2010vd,Gardim:2011xv,Niemi:2012aj,Teaney:2012ke,Qiu:2011iv,Gardim:2014tya,Betz:2016ayq} to result in a strong (nearly) linear mapping between the initial-state eccentricity vectors $\bm{\mathcal{E}_n}$ and the final-state flow vectors $\bm{V_n}$ for both the elliptical ($n=2$) and triangular ($n=3$) harmonics.  We emphasize that this nearly-linear mapping encompasses not just the magnitudes $\varepsilon_n \rightarrow v_n$, but rather the entire complex vectors $\bm{\mathcal{E}_n} \rightarrow \bm{V_n}$.  The strength of this linear mapping motivates a decomposition of the final-state flow vector into a piece which can be predicted directly from the initial-state geometry and a residual:
\begin{align} \label{e:resdef0}
\bm{V_n} \equiv \bm{f} ( \bm{\mathcal{E}_n} ) + \bm{\delta_n}, 
\end{align}
with the vector function $\bm{f}$ of the initial state resulting in the best prediction of the final-state flow when the residuals $\bm{\delta_n}$ are minimized.  A variety of estimator functions have been tested \cite{Gardim:2011xv,Gardim:2014tya,Noronha-Hostler:2015dbi,sievert:2019zjr} to establish the dominance of linear response, with some sensitivity to cubic response as well \cite{Noronha-Hostler:2015dbi,sievert:2019zjr}.  Here we will briefly summarize the general discussion of Ref.~\cite{Noronha-Hostler:2015dbi}, focusing on the contributions of linear + cubic response in practice.

The optimal estimator function $\bm{f} ( \bm{\mathcal{E}_n} )$ is the one for which the expectation value of the residuals
\begin{align} \label{e:minderv0}
\left\langle \delta_n^2  \right\rangle =
\left\langle f_n^2 \right\rangle - 2 \left\langle \mathrm{Re} \left( \bm{V_n} \cdot \bm{f_n} \right) \right\rangle +
\left\langle v_n^2 \right\rangle
\end{align}
is minimized.  Here we use the shorthand $\bm{f_n} \equiv \bm{f} (\bm{\mathcal{E}_n})$ with magnitude $f_n$.  If the estimator function $\bm{f_n}$ depends on a set of parameters $\{ \kappa_i \}$, then the minimization condition $\frac{\partial}{\partial \kappa_i} \langle \delta_n^2 \rangle = 0$ corresponds to
\begin{align} \label{e:minderv1}
0 = \mathrm{Re} \left\langle \left( \bm{V_n} - \bm{f_n} \right) \cdot \frac{\partial}{\partial \kappa_i} \bm{f_n}^* \right\rangle .
\end{align}
In the case of linear + cubic response 
\begin{align} \label{eqn:precubic}
\bm{f} (\bm{\mathcal{E}_n}) = \kappa_{1,n} \bm{\mathcal{E}_n} + \kappa_{2,n} \varepsilon_n^2 \bm{\mathcal{E}_n} ,
\end{align}
the derivatives from \eqref{e:minderv1} lead to
\begin{align} \label{e:minderv2}
0 = \mathrm{Re} \left\langle \left( \bm{V_n} - \bm{f_n} \right) \cdot \bm{f_n}^* \right\rangle = \mathrm{Re} \left\langle \bm{\delta_n} \cdot \bm{f_n}^* \right\rangle.
\end{align}
Eq.~\eqref{e:minderv2} holds not just for linear or linear + cubic response, but for a broad class of estimator functions including polynomials of higher order.

Subject to the optimization condition \eqref{e:minderv2}, the optimized residuals \eqref{e:minderv0} are given by
\begin{align} \label{e:resdef}
\langle \delta_n^2 \rangle = \langle v_n^2 \rangle - \langle f_n^2 \rangle
\end{align}
and can be simply related to the Pearson coefficients $Q_n$:
\begin{align}
\frac{ \langle \delta_n^2 \rangle }{ \langle v_n^2 \rangle } =
1 - Q_n^2,
\end{align}
with the Pearson coefficient being simply a measure of the scalar product between the estimator $\bm{f_n}$ and the resulting flow vector $\bm{V_n}$:
\begin{align}\label{eqn:pear}
Q_n \equiv \frac{\mathrm{Re} \langle \bm{V_n} \cdot \bm{f_n}^* \rangle}{\sqrt{\langle v_n^2\rangle \langle  f_n^2 \rangle}} 
\qquad \overset{\mathrm{linear}}{=} \qquad
\frac{\langle v_n \varepsilon_n \cos\left(n\left[\psi_n-\phi_n\right]\right)\rangle}{\sqrt{\langle \varepsilon_n^2\rangle \langle  v_n^2 \rangle}}.
\end{align}
where the last equality holds only for linear response $\kappa_{2,n} = 0$.  The Pearson coefficient thus quantifies how good the estimator function $\bm{f} (\bm{\mathcal{E}_n})$ is at predicting the flow vector $\bm{V_n}$ by measuring the magnitude of the residuals.  The estimation becomes more accurate in the limit when $| Q_n | \rightarrow 1$, and in the case of linear response, $Q_n \rightarrow +1 (-1)$ reflects a perfect linear (anti-)correlation.

The optimization conditions \eqref{e:minderv1} can also be solved simultaneously to determine the coefficients $\{ \kappa_i \}$ which provide the optimum prediction of the final-state flow vectors.  For the case of linear + cubic response \eqref{eqn:precubic} the result is \cite{Noronha-Hostler:2015dbi}
\begin{subequations}	\label{eqn:nonlinear}
\begin{align}
\kappa_{1,n} &= 
	\frac{\mathrm{Re} \left(
	\langle \varepsilon_n^6\rangle \, 
	\langle \bm{V_n} \cdot \bm{\mathcal{E}_n^*} \rangle - 
	\langle \varepsilon_n^4\rangle \, 
	\langle \bm{V_n} \cdot \bm{\mathcal{E}_n^*} \, \varepsilon_n^2 \rangle\right)
	}
	{
	\langle \varepsilon_n^6\rangle \, \langle \varepsilon_n^2\rangle 
	- \langle \varepsilon_n^4\rangle^2
	} ,
\\
\kappa_{2,n} &= 
	\frac{\mathrm{Re} \left(
	-\langle \varepsilon_n^4 \rangle \, 
	\langle \bm{V_n} \cdot \bm{\mathcal{E}_n^*}\rangle + 
	\langle \varepsilon_n^2\rangle \, 
	\langle \bm{V_n} \cdot \bm{\mathcal{E}_n^*} \, \varepsilon_n^2 \rangle\right)
	}
	{
	\langle \varepsilon_n^6\rangle \, 
	\langle \varepsilon_n^2\rangle - \langle \varepsilon_n^4\rangle^2
	} .
\end{align} 
\end{subequations}
Finally, we note that the optimization condition \eqref{e:minderv2} can be used to decompose the two- and four-particle cumulants in terms of the estimator and residuals:  For the two-particle cumulant we have
\begin{subequations} \label{e:optcum1}
\begin{align}
(v_n\{2\})^2 &\equiv \langle v_n^2 \rangle
\notag \\ &=
\langle | \bm{f_n} + \bm{\delta_n} |^2 \rangle
\notag \\ &=
\langle f_n^2 \rangle + \langle \delta_n^2 \rangle
\\ \notag \\
(v_n \{4\})^4 &\equiv 2 \langle v_n^2 \rangle^2 - \langle v_n^4 \rangle
\notag \\ &=
2 \langle | \bm{f_n} + \bm{\delta_n} |^2 \rangle^2 - 
\left[
\langle f_n^4 \rangle + 
\left( \langle v_n^4 \rangle - \langle f_n^4 \rangle \right)
\right]
\notag \\ &=
2 \left( \langle f_n^2 \rangle + \langle \delta_n^2 \rangle \right) ^2 - \langle f_n^4 \rangle - \Delta_{n,4}
\end{align}
\end{subequations}
where we have used the condition \eqref{e:minderv2} and defined a new quantity
\begin{eqnarray}	\label{eqn:d4}
\Delta_{n,4} &\equiv& \langle v_n^4 \rangle - \langle f_n^4 \rangle.
\end{eqnarray}
We note that $\Delta_{n,4} \neq \langle \delta_n^4 \rangle$, differing by mixed terms which are not needed for our calculation.  For the case of linear + cubic response \eqref{eqn:precubic}, Eqs.~\eqref{e:optcum1} can be written
\begin{subequations}	\label{e:optcum2}
\begin{align}
v_n \{2\} &= \sqrt{\left\langle \big(\kappa_{1,n} \varepsilon_n+\kappa_{2,n} \varepsilon_n^3\big)^2\right\rangle + \big\langle \delta_n^2 \big\rangle}
\\ \notag \\
\frac{v_n\{4\}}{v_n\{2\}} &= 
	\frac{\sqrt[4]{
	2\left( 
		\left\langle \Big(\kappa_{1,n} \varepsilon_n + 
		\kappa_{2,n} \varepsilon_n^3 \Big)^2 \right\rangle 
		+ \big\langle \delta_n^2 \big\rangle 
	\right)^2 
	- \left\langle \big(\kappa_{1,n} \varepsilon_n + \kappa_{2,n} \varepsilon_n^3 \big)^4 \right\rangle - \Delta_{n,4}
	}}
	{
	\sqrt{\left\langle \big(\kappa_{1,n} \varepsilon_n+\kappa_{2,n} \varepsilon_n^3\big)^2\right\rangle + \big\langle \delta_n^2 \big\rangle}
	} ,
\end{align}
\end{subequations}
where we emphasize the ratio $\frac{v_n\{4\}}{v_n\{2\}}$ which is directly sensitive to the fluctuations of the flow vector $\bm{V_n}$.

In this manner we have decomposed the contributions to the final flow harmonics into initial state effects (the eccentricities $\varepsilon_n$), final state effects (the response parameters $\kappa_{1,n}$ and $\kappa_{2,n}$), and the residuals $\delta_n$ and $\Delta_{n,4}$.  The contributions $\varepsilon_n$ from the fluctuating initial state are made explicit, while the response coefficients encapsulate various medium effects such as the lifetime of the hydrodynamic phase, transport coefficients and the equation of state.  Additionally, the residuals $\delta_n$ and $\Delta_{n,4}$ represent a mixture of both initial and final state effects, encoding the remaining contributions of unknown origin to the observed flow.  The particular residuals $\delta_n$ and $\Delta_{n,4}$ calculated here are specific to linear + cubic response; a different choice of the estimator function $\bm{f}$ could move some part of these residuals into additional explicit initial and final state factors.  In Sec.\ \ref{sec:kappas} we will extract $\kappa_{1,n}, \kappa_{2,n}, \delta_n$, and $\Delta_{n,4}$ from the top three beam energies and extrapolate downwards to lower beam energies in order to make baseline predictions for $v_n\{2\}$ and $\frac{v_n\{4\}}{v_n\{2\}}$ if the high-energy paradigm were to remain unmodified.

%
\subsection{Comparison of Trento and MC-KLN Initial Conditions}
\label{sec:TrentoMCKLN}
%

One of the most stringent constraints for initial condition models is the necessity to match the event-by-event fluctuations of $v_2$ \cite{Giacalone:2017uqx}, which are characterized by the ratio of the four- and two-particle cumulants:
\begin{align} \label{e:cumratio}
\frac{v_n \{4\}}{v_n \{2\}} = \sqrt[4]{1 - \frac{\mathrm{Var}(v_n^2)}{\langle v_n^2 \rangle^2}} .
\end{align}
As seen in Eq.~\eqref{e:cumratio}, when the ratio $v_2\{4\}/v_2\{2\}$ approaches 1 there are fewer fluctuations in the system, whereas when $v_2\{4\}/v_2\{2\} \ll 1$ there are more fluctuations in the system.  In central collisions at top beam energies, linear response dominates and to a good approximation we have
\begin{equation}
\frac{v_n\{4\}}{v_n\{2\}} \approx \frac{\varepsilon_n\{4\}}{\varepsilon_n\{2\}} ,
\end{equation}
with the coefficient $\kappa_{1,n}$ canceling in the ratio.  In mid-central collisions at top energies, linear+cubic response dominates and one can still predict the cumulant ratio $v_2\{4\}/v_2\{2\}$ reasonably well using only $\kappa_{1,2}$ and $\kappa_{2,2}$ \cite{sievert:2019zjr}. 

\begin{figure}[h]
	\centering
	\begin{tabular}{c c}
		\includegraphics[width=0.5\linewidth]{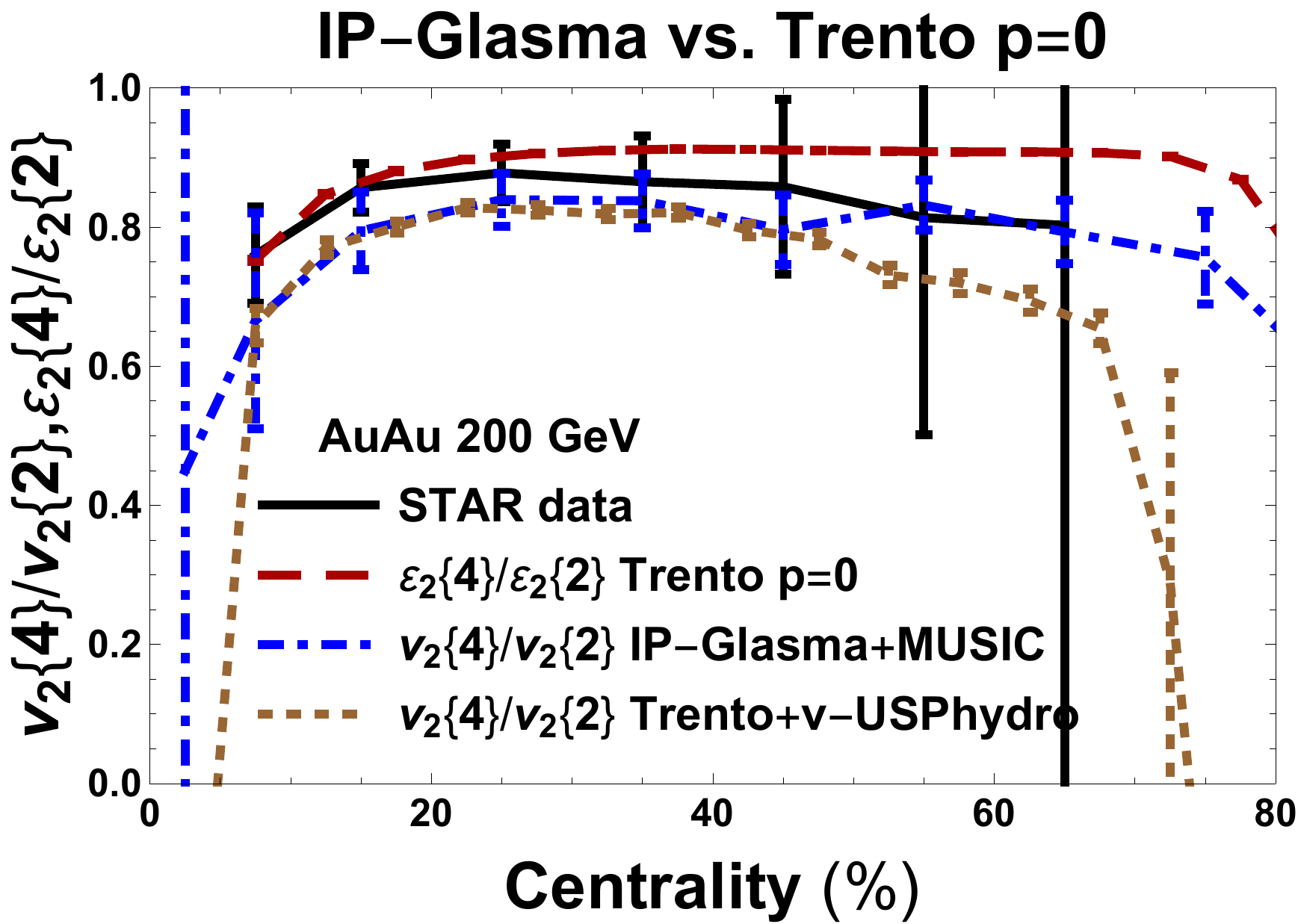} & \includegraphics[width=0.5\linewidth]{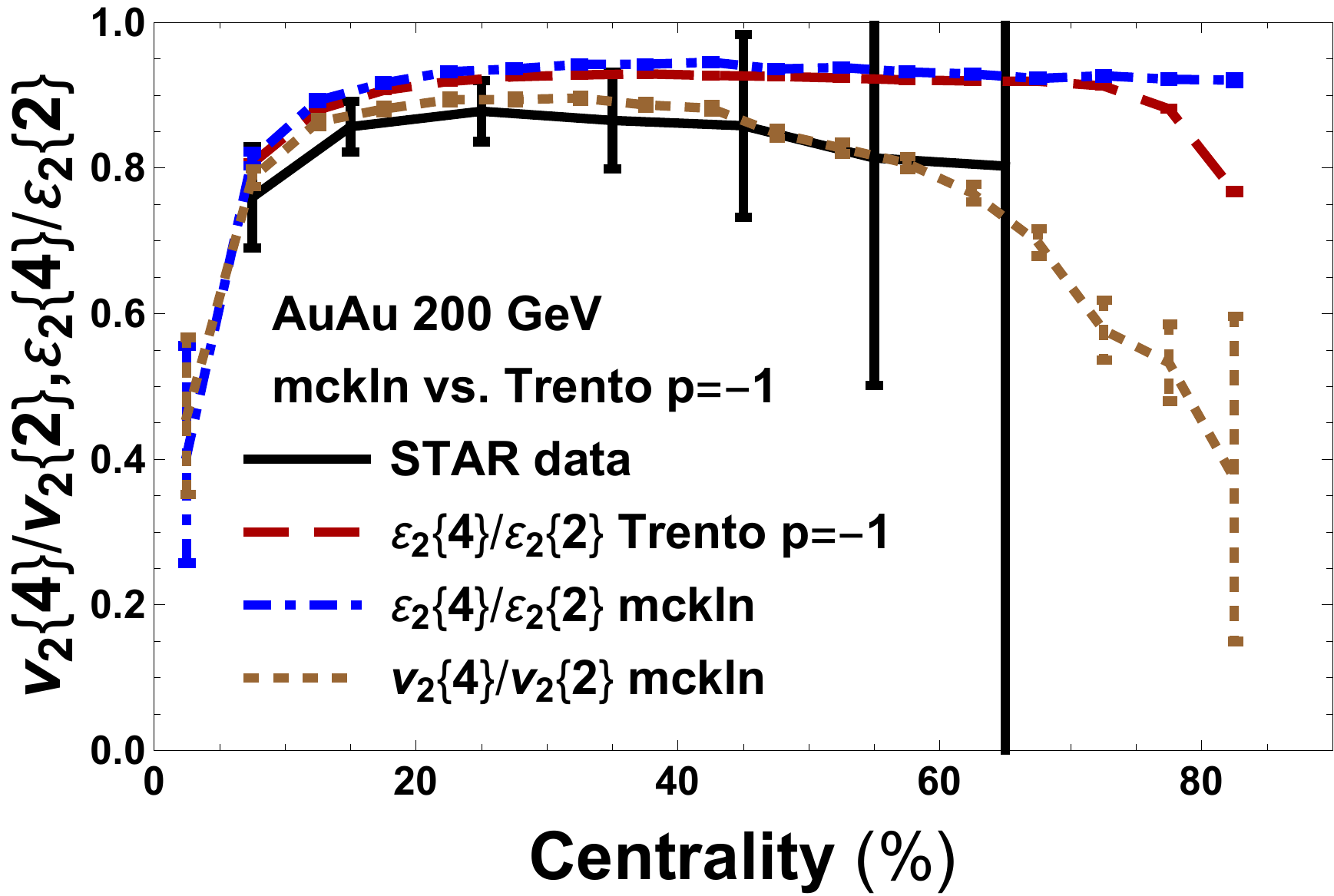} 
	\end{tabular}
	\caption{(Color online) Comparison of $v_2\{4\}/v_2\{2\}$ STAR data to TRENTO and MC-KLN eccentricities and hydro calculations from Trento+v-USPhydro, MC-KLN+v-USPhydro, and IP-Glasma+MUSIC.  STAR data is taken from \cite{Adams:2004bi} and the IP-Glasma+MUSIC data is from \cite{Schenke:2019ruo}.}
	\label{fig:datcomp}
\end{figure}

As a starting point, let us compare three prominent models (Trento\footnote{When we specify Trento we assume the default parameters unless specified elsewhere}, IP-Glasma, and MC-KLN) for the initial conditions at top collider energies both before and after hydrodynamic evolution.  As pointed out in Ref.~\cite{Katz:2019fkc}, Trento+v-USPhydro works best when compared to the highest LHC beam energies, while MC-KLN does not fluctuate enough on an event-by-event basis to adequately describe the cumulant ratio $v_2\{4\}/v_2\{2\}$ \cite{Drescher:2006ca}.  However, both initial-state models give relatively comparable results for the RMS flow measured by the two-particle cumulants $v_2\{2\}$ and $v_3\{2\}$, and the fluctuations in MC-KLN do provide a slightly better fit to RHIC data for $v_2\{4\}/v_2\{2\}$.  In the case of MC-KLN+v-USPhydro, $\eta/s$ must change significantly more with beam energy to describe the data, such that $\eta/s=0.05$ for PbPb 5.02 TeV, $\eta/s=0.11$ for PbPb 2.76 TeV, and $\eta/s=0.08$ for AuAu 200 GeV.  These MC-KLN+v-USPhydro simulations use the same initialization time $\tau_0=0.6$ fm and a lower freeze-out temperature of $T_{FO}=120$ MeV compared to Trento+v-USPhydro. Additionally, as previously mentioned the MC-KLN+v-USPhydro are based on an outdated equation of state so they are used only as a comparison to Trento+v-USPhydro.

In the left panel of Fig.\ \ref{fig:datcomp} we compare the STAR data \cite{Adams:2004bi} on AuAu collisions at 200 GeV to both eccentricities and full hydrodynamic calculations using either the ``IP-Glasma-like'' configuration of Trento ($p=0$) or actual IP-Glasma initial conditions.  In central collisions (which are dominated by initial state effects), we find that the results of the hydrodynamic calculations Trento+v-USPhydro and IP-Glasma+MUSIC  are essentially equivalent.  In very peripheral collisions the predictions begin to deviate, which is expected since the two hydrodynamic models have very different assumptions and parameters.  Additionally, we compare these same results to the initial-state eccentricities $\varepsilon_2\{4\}/\varepsilon_2\{2\}$ calculated in Trento $p=0$, which also agree well with the data.

In the right panel of Fig.\ \ref{fig:datcomp} we then compare the MC-KLN scenario.  In Ref.~\cite{Bernhard:2016tnd}, it was discussed that Trento with the setting $p=-0.67$ gives roughly equivalent initial eccentricities to MC-KLN.  Here we compare the eccentricity fluctuations $\varepsilon_2\{4\}/\varepsilon_2\{2\}$ produced by MC-KLN versus the Trento setting $p=-1$.  Despite this being a more extreme value than the $p=-0.67$ value preferred in Ref.~\cite{Bernhard:2016tnd}, we still find that Trento $p=-1$ produces nearly identical eccentricity fluctuations to MC-KLN.  We conclude that we can replicate well the initial state produced with MC-KLN by instead running Trento with $p=-1$ (and we prefer a more extreme value to test a wider range of possibilities at the lowest beam energy).  Previous calculations using the full hydrodynamic evolution of MC-KLN+v-USPhydro compare favorably to the STAR data, and we further note that the initial-state eccentricities $\varepsilon_2\{4\}/\varepsilon_2\{2\}$ in either model closely track the final flow measurements in central collisions.

Overall, at RHIC energies Trento $p=0$, IP-Glasma, and MC-KLN all provide a reasonable description of the experimental data.  Additionally, we find that Trento with settings $p=0$ and $p=-1$ seems to be a good proxy for IP-Glasma and MC-KLN initial conditions, respectively.  Thus, in Sec.\ \ref{sec:initialstate} on eccentricities we will compare the different values of $p$ in Trento to explore the behavior of different initial condition  models at lower beam energies.

%
\section{Beam energy dependence of $\varepsilon_2$ fluctuations}
\label{sec:initialstate}
%

As a first step, we study the beam energy dependence arising from the initial conditions alone, as quantified by the eccentricity cumulant $\varepsilon_2\{2\}$ and cumulant ratio $\varepsilon_2\{4\}/\varepsilon_2\{2\}$.  We calculate these quantities using Trento, but varying the parameter $p$ which controls the determination of the reduced thickness function $T_R$ between $p=-1$ (MC-KLN-like), $p=0$ (IP-Glasma- / EKRT-like), and $p=1$ (Glauber-like).  We also consider possible modifications at AuAu $\sqrt{s_{NN}}=7$ GeV to the other parameters in Trento, such as the nucleon width $\sigma$ and parameter $k$ controlling the tails of the multiplicity distribution.

\begin{table}[h]
	\begin{tabular}{|c|c|}
		\hline
		$\sqrt{s_{NN}}$ [GeV] & $\sigma^{inel}_{NN}$ [mb]\\
		\hline
		200 & 42.3 \\
		54 & 35 \\
		27 & 33.2 \\
		7.7 & 31.2 \\
		\hline
	\end{tabular}
	\caption{Table of inelastic cross-sections used to calculate the eccentricities in Trento for the Beam Energy Scan energies.  All are taken from Ref.~\cite{Adare:2015bua} except for $\sqrt{s_{NN}}$=54 GeV, which was set by estimating using surrounding beam energies.}
	\label{tab:cross}
\end{table}

In Trento, the energy dependence of the initial conditions arises solely from setting the experimentally-measured nucleon-nucleon inelastic cross section $\sigma_{NN}^{inel}$ \cite{Moreland:2014oya}.  This experimental input is used to indirectly tune the effective partonic scattering cross section $\sigma_{gg}$ which enters the collision probability $P_\mathrm{coll} = 1 - \exp\left[-\sigma_{gg} \int d^3 x \, \rho_A (\vec{x}) \, \rho_B (\vec{x}) \right]$ between two nucleons.  As seen in Table \ref{tab:cross}, this energy dependence is particularly mild, decreasing only $\sim\ord{25\%}$ over two orders of magnitude in $\sqrt{s_{NN}}$.  The success of the mild energy dependence implemented in Trento in describing the initial conditions from top RHIC to LHC energies is attributable to the fact that high-energy cross sections in QCD are energy independent at leading order.  In the high-energy (``eikonal'') limit $s \rightarrow \infty$ of QCD, a scattering cross section $\sigma$ can be expanded in powers of the energy $s$ as
\begin{align} \label{e:s_expansion}
\sigma = \left(\tfrac{\mu^2}{s}\right)^0 \sigma_\mathrm{eik}
+ \left(\tfrac{\mu^2}{s}\right)^1 \sigma_\mathrm{sub-eik} + \cdots
\end{align}
at leading order, with $\mu^2$ some fixed transverse scale to make the expansion parameter dimensionless.  This hierarchy of power-suppressed terms is accurate at leading order, with certain higher orders in $\alpha_s$ generating additional logarithmic dependence on the energy through powers of $\ln\tfrac{s}{\mu^2}$.  A systematic resummation of such logarithmic terms leads to a small enhancement in the overall power of $(\mu^2 / s)$, but it does not overturn the leading-order decrease of $\sigma_\mathrm{sub-eik}$ with energy \cite{Kirschner:1983di, Griffiths:1999dj, Itakura:2003jp, Kovchegov:2015pbl, Kovchegov:2016zex}.

%
\begin{figure}
	\includegraphics[width=0.5\textwidth]{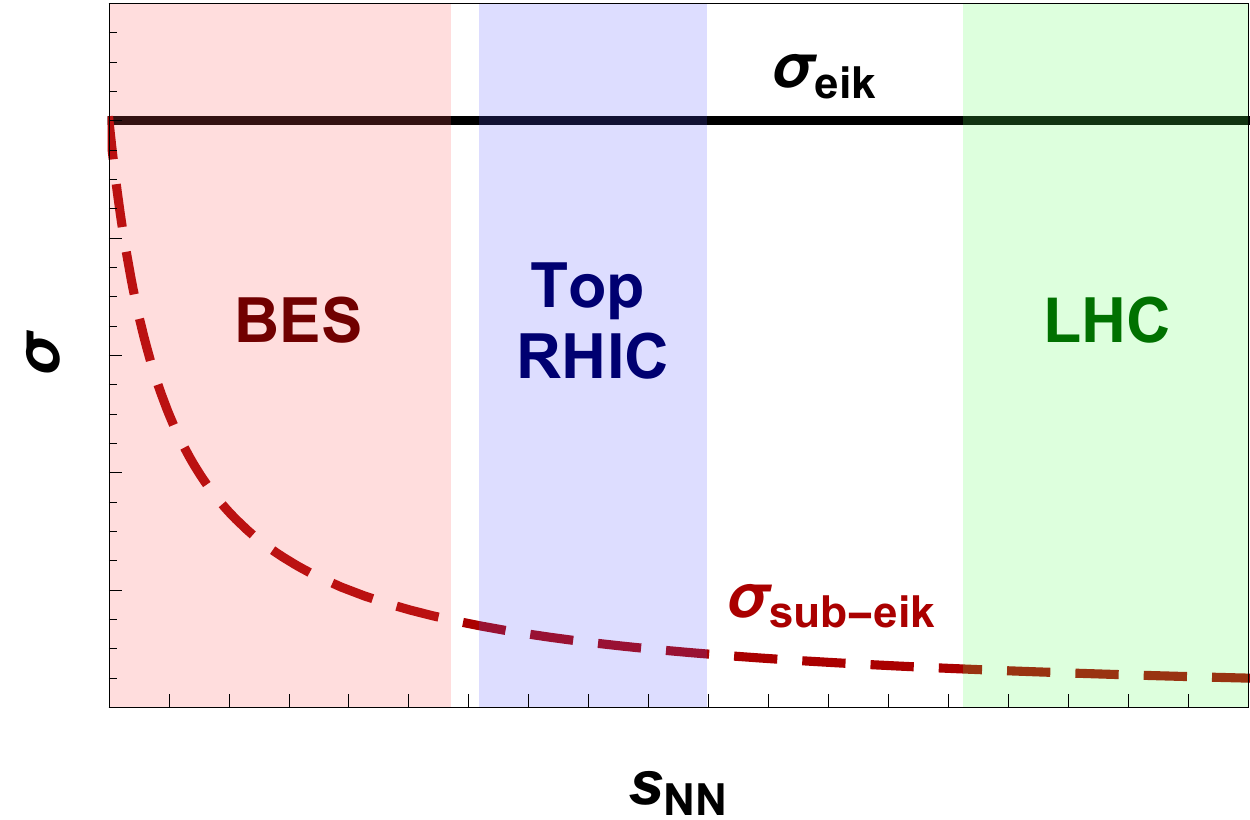}
	\caption{Cartoon plot of Eq.~\eqref{e:s_expansion} illustrating the comparison between the energy-independent cross section $\sigma_\mathrm{eik}$ and the energy-suppressed corrections $\sigma_\mathrm{sub-eik}$ at high energies.  Shaded regions sketch the kinematic regions of various collider programs, illustrating the dominance of $\sigma_\mathrm{eik}$ at LHC and top RHIC energies in contrast to the Beam Energy Scan (BES) where the new mechanisms contained within $\sigma_\mathrm{sub-eik}$ can become very important.
	} 
	\label{f:xsecplot}
\end{figure}
%

The leading term $\sigma_\mathrm{eik}$ is energy independent and thus survives in the eikonal approximation $s \rightarrow \infty$.  The independence of $\sigma_\mathrm{eik}$ from the collision energy is equivalent to independence with respect to the total rapidity interval $\Delta Y \sim\ln\tfrac{s_{NN}}{m_N^2}$ and thus to {\it{boost invariance}} of the initial state.  As is well known, at high energy (synonymous with small $x$) the initial state corresponding to $\sigma_\mathrm{eik}$ is dominated by abundant soft gluon radiation which constitutes the initial energy density of a heavy-ion collision.  The physics of baryon stopping \cite{Itakura:2003jp}, along with spin dependence \cite{Kovchegov:2015pbl}, medium-induced radiation \cite{Sievert:2018imd}, and many other effects are power suppressed at high energies, belonging to the sub-eikonal cross section $\sigma_\mathrm{sub-eik}$ or higher-order terms.  As illustrated in Fig.~\ref{f:xsecplot}, these contributions die off at top collider energies to yield the well-known gluon-dominated initial state which is implemented in the various models.  But as the energy is {\it{lowered}}, the neglect of these sub-eikonal effects becomes a poorer and poorer approximation.  

All of the models considered here restrict themselves in some fashion to the gluon-dominated physics contained in $\sigma_\mathrm{eik}$, so naturally all of them will lead to a very mild energy dependence.  But this weak energy dependence of the initial state predicted in the various models is in some sense artificial: a consequence of being tuned to the (approximately) energy-independent initial conditions relevant for top collider energies.  The initial state {\it{does}} and {\it{must}} deviate from these predictions in significant ways at lower energies, as the sub-eikonal physics (including baryon stopping in particular) becomes increasingly important.  The purpose here is to extrapolate these models down to lower energies to identify where and how the deviations from eikonally-produced initial conditions occurs.  We also note that the different models incorporate the explicit energy dependence differently, so matching Trento $p=0$ and $p=-1$ to other models like IP-Glasma and MC-KLN as in Sec.~\ref{sec:TrentoMCKLN} and then extrapolating downward in energy with Trento is {\it{not}} the same as directly extrapolating these original models themselves.  For this reason, we will carefully refer to our initial condition models as $p=0$ (IP-Glasma-like) and $p=-1$ (MC-KLN-like), rather than to the actual IP-Glasma or MC-KLN models as appropriate.

%
\begin{figure}[h]
	\centering
	\begin{tabular}{c c}
		\includegraphics[width=0.5\linewidth]{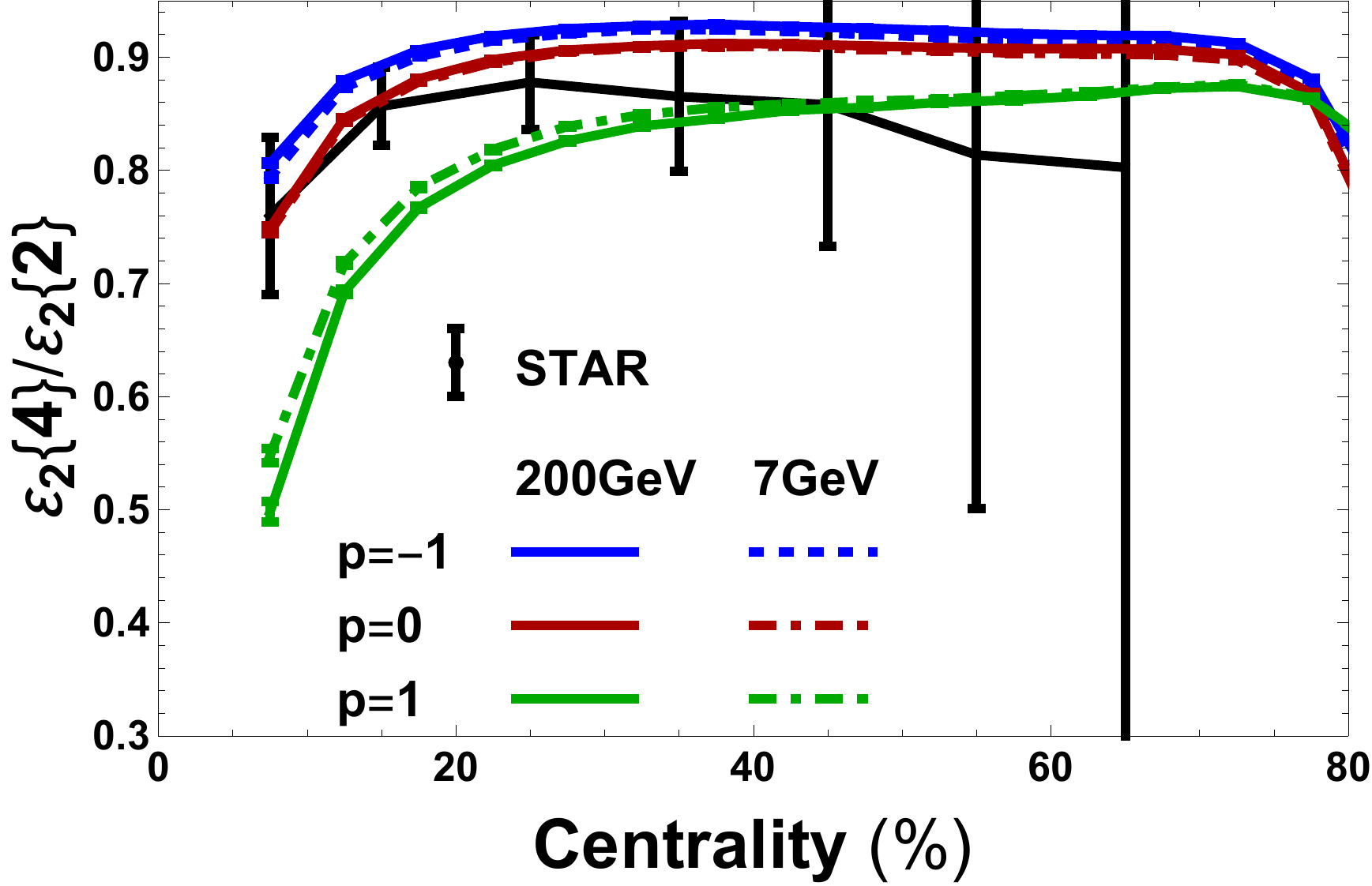} & \includegraphics[width=0.5\linewidth]{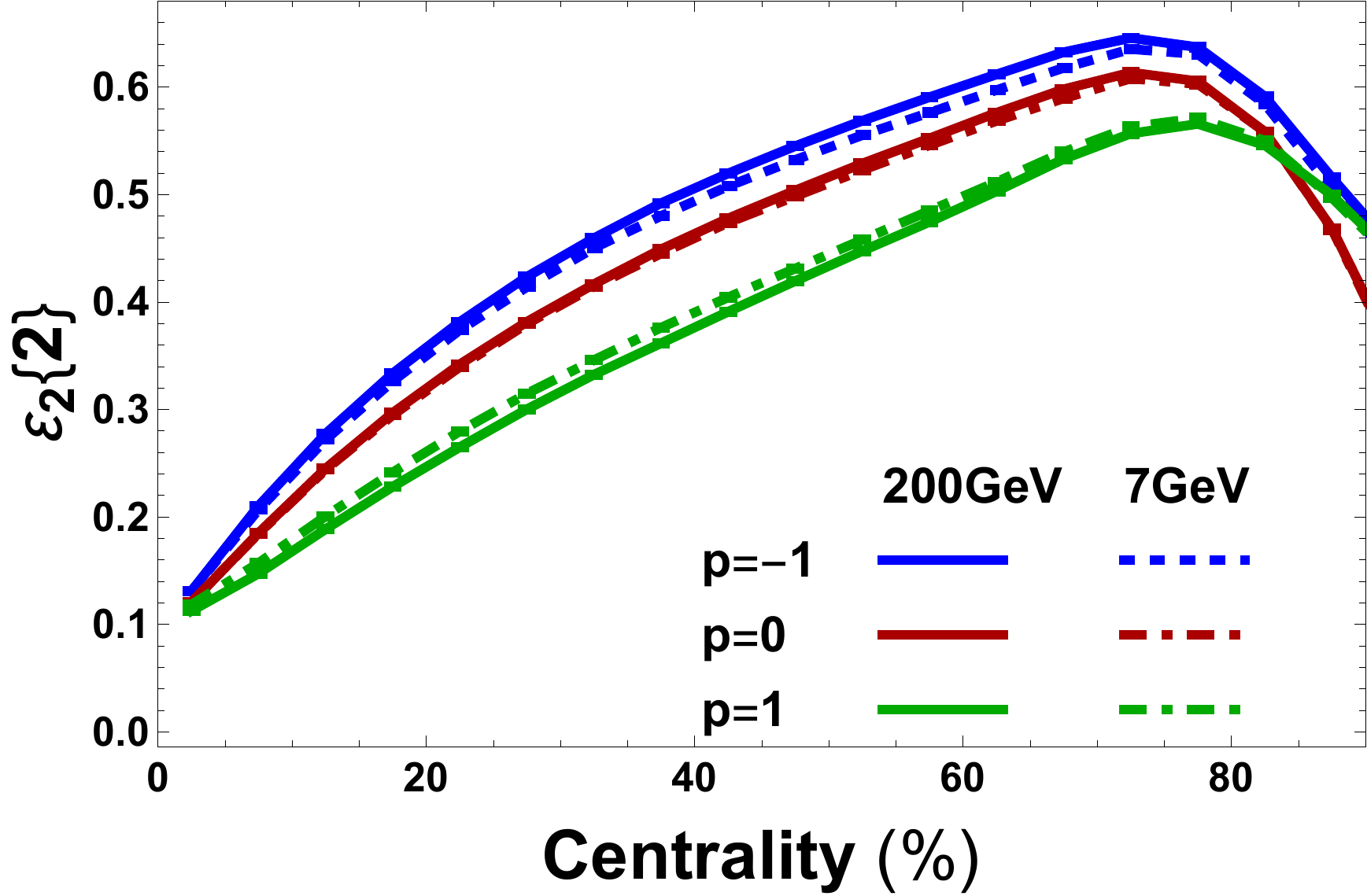} 
	\end{tabular}
	\caption{(Color online) Comparison of $\varepsilon_2\{4\}/\varepsilon_2\{2\}$ (left) and $\varepsilon_2\{2\}$ (right) for three descriptions of the initial state in Trento:  $p=-1$ (MC-KLN-like), $p=0$ (IP-Glasma/EKRT-like), and $p=1$ (Glauber-like).  The $200 \mathrm{GeV}$ AuAu STAR data from Ref.~\cite{Adamczyk:2015obl} is shown in black.  For the initial state models, two different beam energies of AuAu $\sqrt{s_{NN}}=200$ GeV and  $\sqrt{s_{NN}}=7$ GeV are considered.}
	\label{fig:eccps}
\end{figure}
%

In Fig.\ \ref{fig:eccps} we plot the three scenarios: Trento $p=-1$ (MC-KLN-like), Trento $p=0$ (IP-Glasma/EKRT-like), and Trento $p=1$ (Glauber-like).  In Ref.~\cite{Giacalone:2017uqx} it was shown that the comparison of $v_2\{4\}/v_2\{2\}$ in ultracentral collisions could disfavor the Glauber model at LHC energies.  Here we find that even at RHIC $\sqrt{s_{NN}}=200$ GeV the Glauber-like model ($p=1$) is disfavored, and we are unaware of any final state effects that could correct such a significant initial-state disparity.  As discussed previously, both MC-KLN and IP-Glasma provide a reasonable comparison of the STAR data \cite{Adamczyk:2015obl} at 200 GeV.  And for the reasons we have anticipated above, we see that there is essentially no beam energy dependence of $\varepsilon_2\{4\}/\varepsilon_2\{2\}$ and $\varepsilon_2\{2\}$.  Curiously, the Glauber-like initial conditions ($p=1$) appear to have a slight increase in $\varepsilon_2\{4\}/\varepsilon_2\{2\}$ at $\sqrt{s_{NN}}=7 \mathrm{GeV}$, whereas $p=0$ and $p=-1$ are slightly suppressed at lower beam energies.  The magnitudes of  $\varepsilon_2\{2\}$ are also nearly identical across beam energies, with the MC-KLN-like setting $p=-1$ producing the largest eccentricities, followed by the IP-Glasma-like $p=0$ and then the Glauber-like $p=1$.  From the eccentricities shown here, extrapolated from the eikonal models, one would anticipate no significant changes in $v_2\{4\}/v_2\{2\}$ and $v_2\{2\}$ due to the initial state when the beam energies are decreased.  These results are consistent with  \cite{Noronha-Hostler:2015uye} where the eccentricities from LHC run 1 and run 2 were compared, although only the 2 particle correlation was considered in that paper.  We note that the STAR data for $v_2\{4\}/v_2\{2\}$ shown in Figs.\ \ref{fig:datcomp}-\ref{fig:eccps} is obtained from measurements of $v_2\{2\}$ and $v_2\{4\}$ reported separately, rather than reporting the ratio directly.  Unfortunately, this requires us to use error propagation to combine the uncertainties in the two measurements; since these errors are correlated, this method likely overestimates the true experimental errors.  For this reason, we urge experimentalists to always report directly the ratio and its uncertainties as well as the individual measurements. 

%
\begin{figure}[h]
\centering
\includegraphics[width=0.5\linewidth]{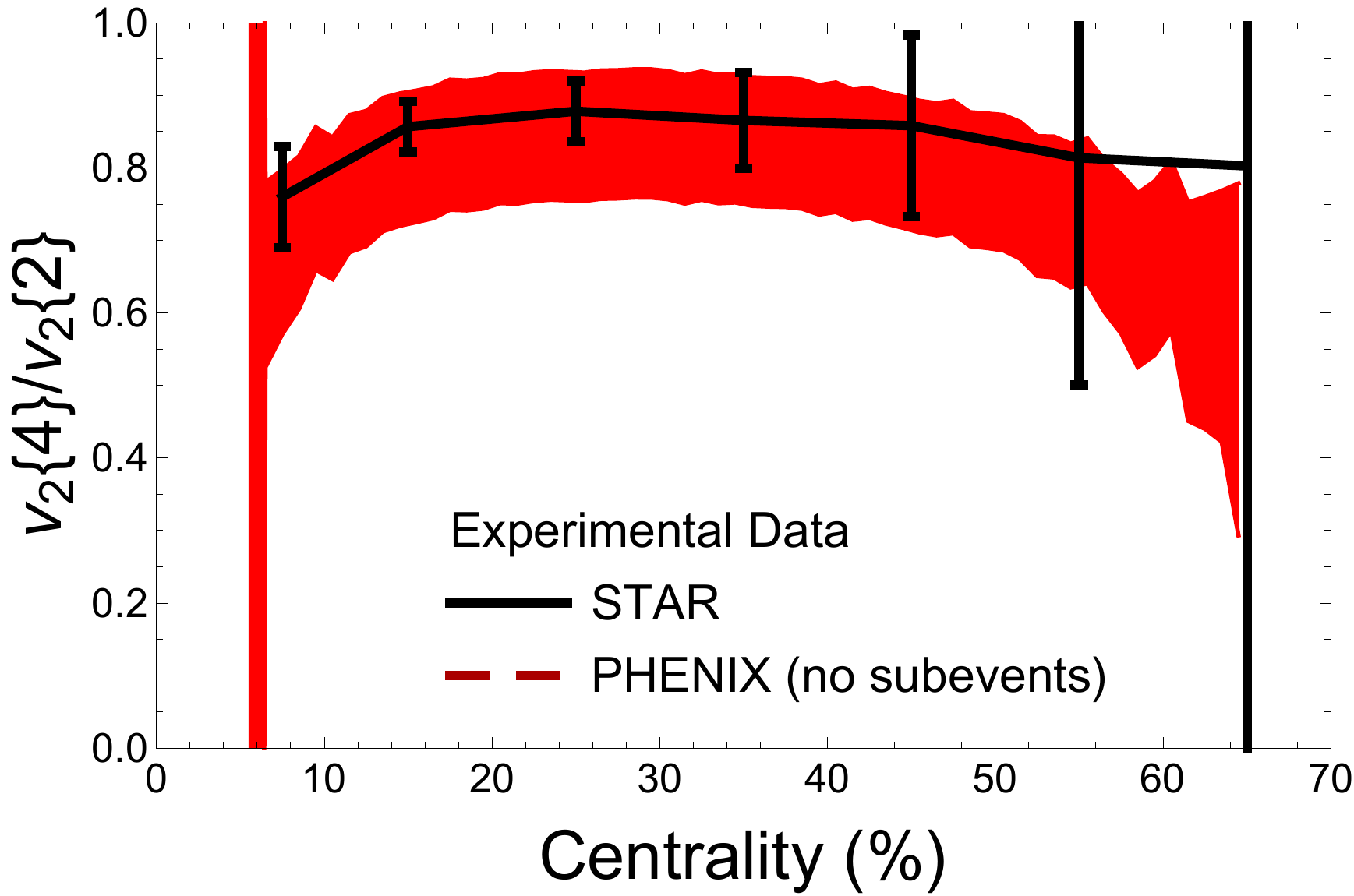} 
\caption{(Color online) Comparison of STAR and PHENIX experimental data for  $v_2\{4\}/v_2\{2\}$ at RHIC AuAu $\sqrt{s_{NN}}=200$ GeV.}
\label{fig:STARPHENIX}
\end{figure}
%

PHENIX has also recently published new data on $v_2\{2\}$ and  $v_2\{4\}$, which we plot in Fig.\ \ref{fig:STARPHENIX} and compare to the aforementioned STAR data by again taking the ratio $v_2\{4\}/v_2\{2\}$ and propagating the individually-determined error bars.  In this data, where no subevents are considered, the PHENIX results are consistent with the STAR data.  However PHENIX has also calculated $v_2\{4\}/v_2\{2\}$ in finer centrality bins, so the error bars are larger (the exception being the most peripheral collisions).  Since we are primarily concerned with central to mid-central collisions in this paper, for this particular observable, we will focus on comparisons to the STAR data throughout the rest of this paper.  

%
\begin{figure}[h]
\centering
\includegraphics[width=0.5\linewidth]{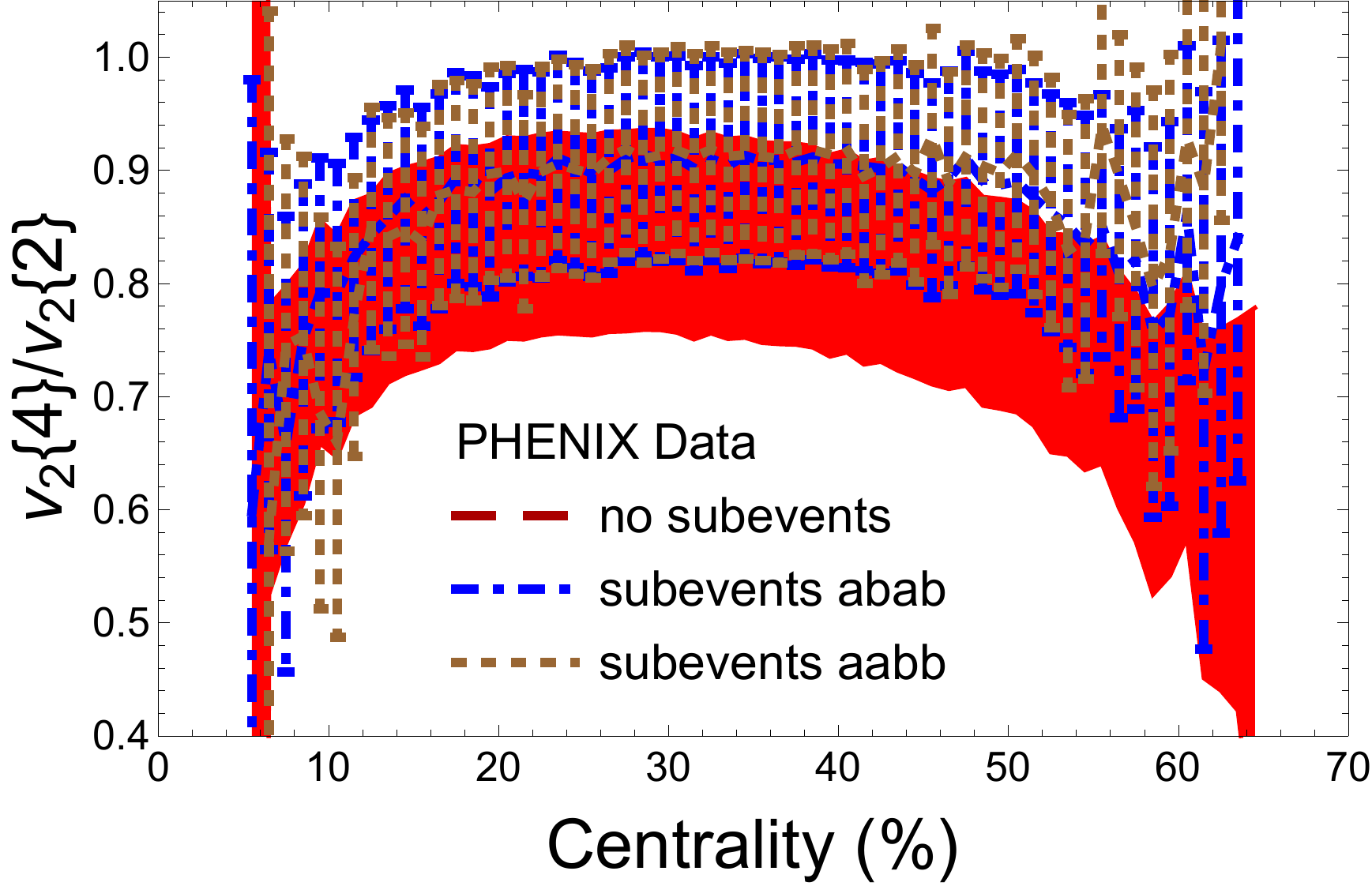} 
\caption{(Color online) Comparison of PHENIX experimental data with and without subevents for  $v_2\{4\}/v_2\{2\}$ at RHIC AuAu $\sqrt{s_{NN}}=200$ GeV.}
\label{fig:subevents}
\end{figure}
%

We do, however, note that there are significant differences if subevents are considered, which could significantly impact our choice in initial conditions.  In Fig.\ \ref{fig:subevents} the red band is the PHENIX data for $v_2\{4\}/v_2\{2\}$ without selecting on subevents, as in Fig.\ \ref{fig:STARPHENIX}.  Once subevents are included, the ratio $v_2\{4\}/v_2\{2\}$ moves substantially toward unity, although there is still sizable overlap in the error bars.  Thus, it would be very interesting to re-bin the subevent data into larger centrality bins to reduce the size of the error bars in order to further discriminate between initial conditions.  This, however, is a task we must leave to the experimentalists. 

%
\begin{figure}[h]
	\centering
	\begin{tabular}{c c}
		\includegraphics[width=0.5\linewidth]{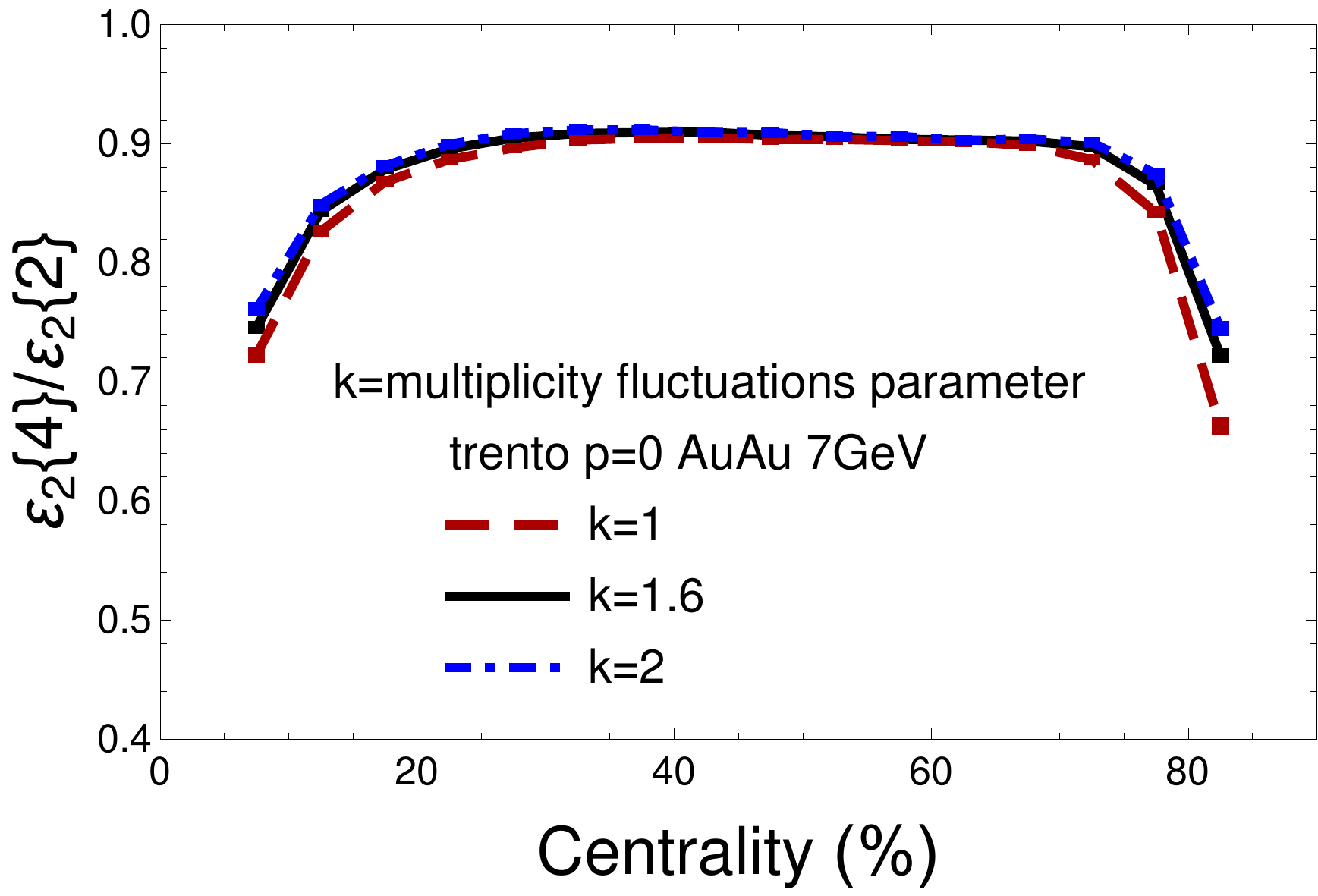} & \includegraphics[width=0.5\linewidth]{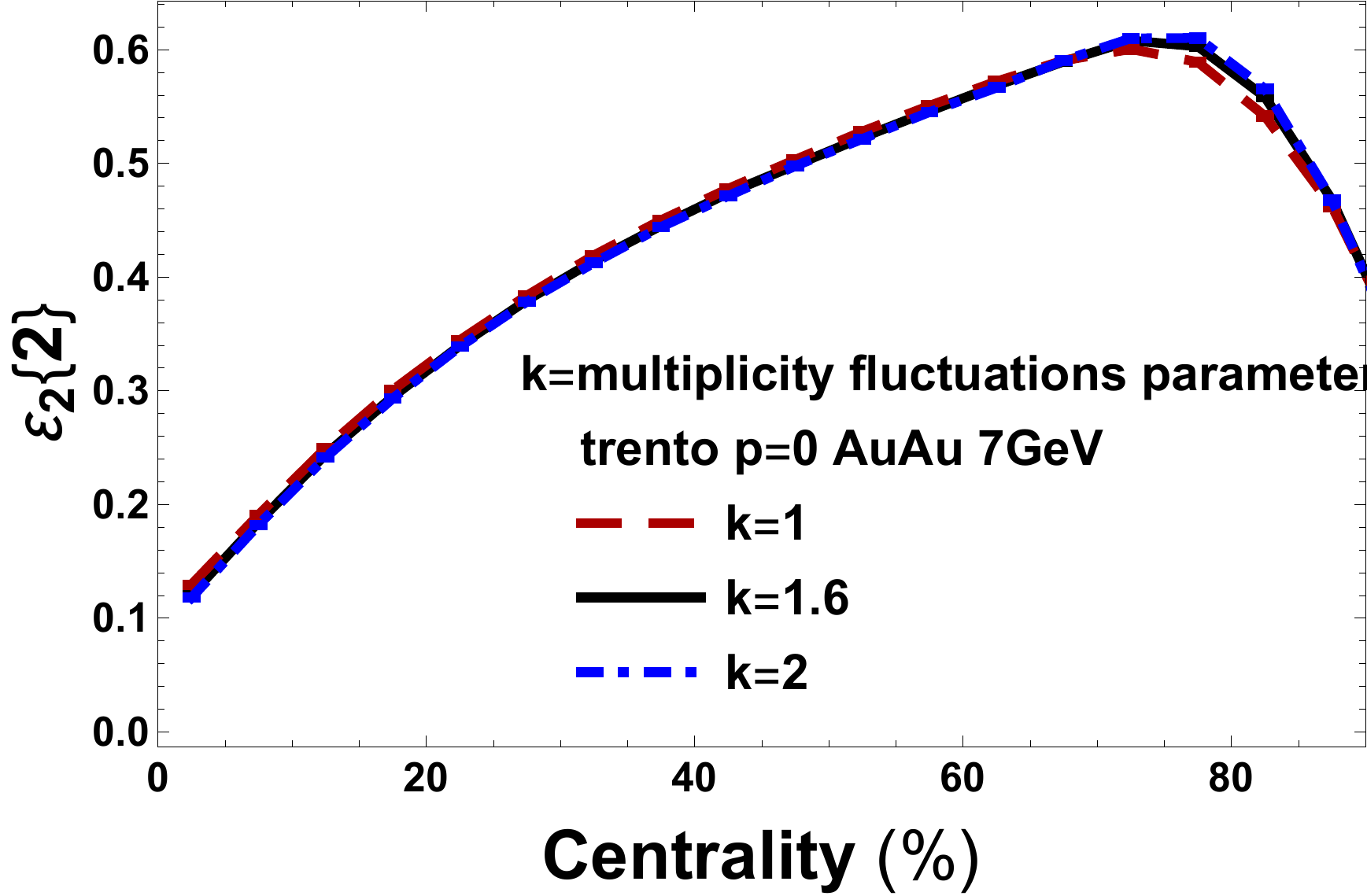} 
	\end{tabular}
	\caption{(Color online) Comparison of $\varepsilon_2\{4\}/\varepsilon_2\{2\}$ (left) and $\varepsilon_\{2\}$ (right) for varied k fluctuations value in Trento where k=1.6 is the standard value from Bayesian analysis.  Calculations done at AuAu  $\sqrt{s_{NN}}=7$ GeV.}
	\label{fig:kpar}
\end{figure}
%

In addition to the different $p$ values mimicking different initial condition models, there are still other changes which may occur at lower beam energies beyond the dependence incorporated by changing the nucleon-nucleon cross section.  For instance, multiplicity fluctuations\footnote{Note that here we are not discussing net-baryon fluctuations, which arise from entirely different physics.}
may change as one decreases the beam energy.  Thus in Fig.\ \ref{fig:kpar} we also investigate the influence of the Trento parameter $k$ which drives the multiplicity fluctuations.  Generally, we don't see a strong dependence on $k$ at $\sqrt{s_{NN}}=7$ GeV if $k$ were to change with beam energy.

%
\begin{figure}[h]
	\centering
	\begin{tabular}{c c}
		\includegraphics[width=0.5\linewidth]{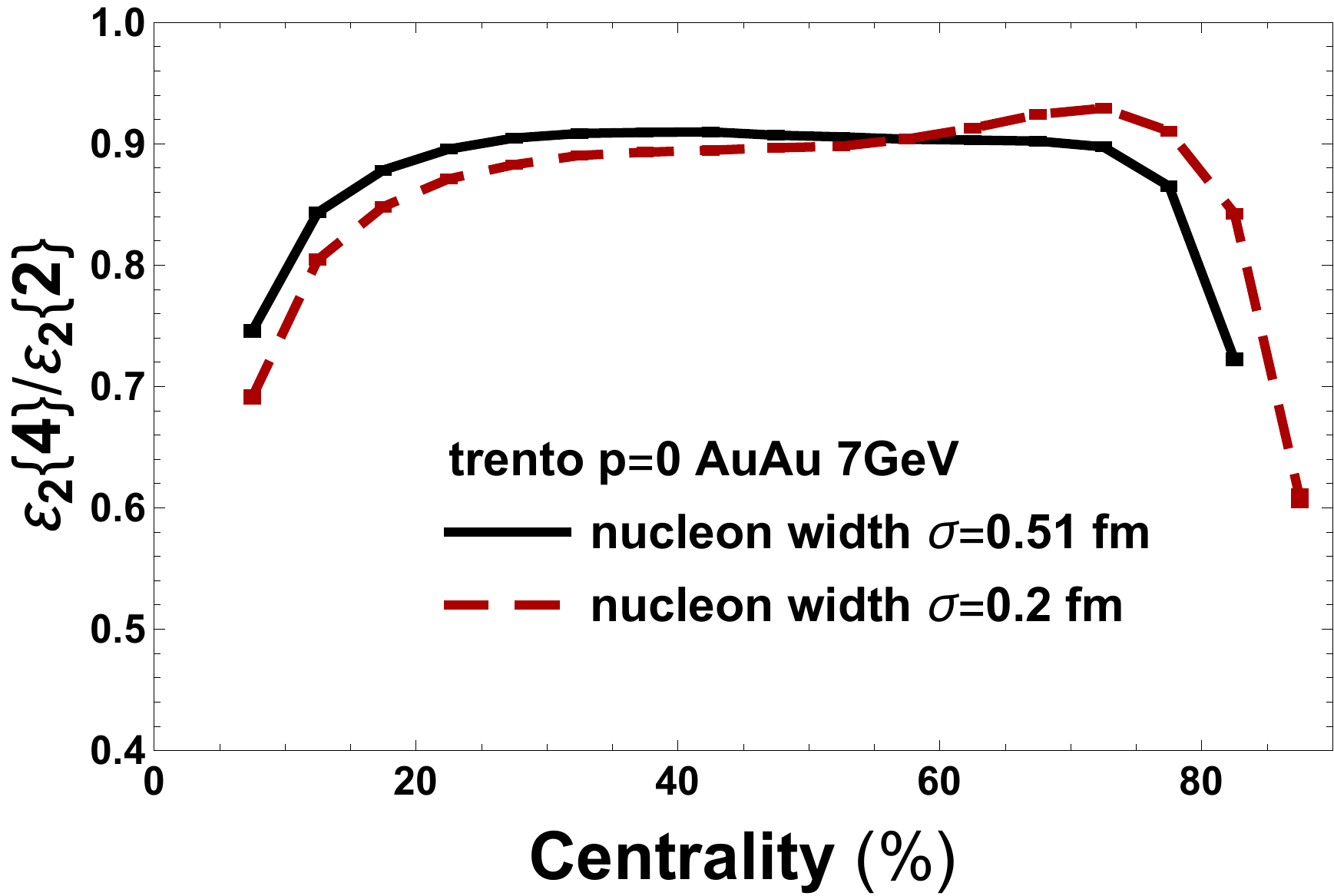} & \includegraphics[width=0.5\linewidth]{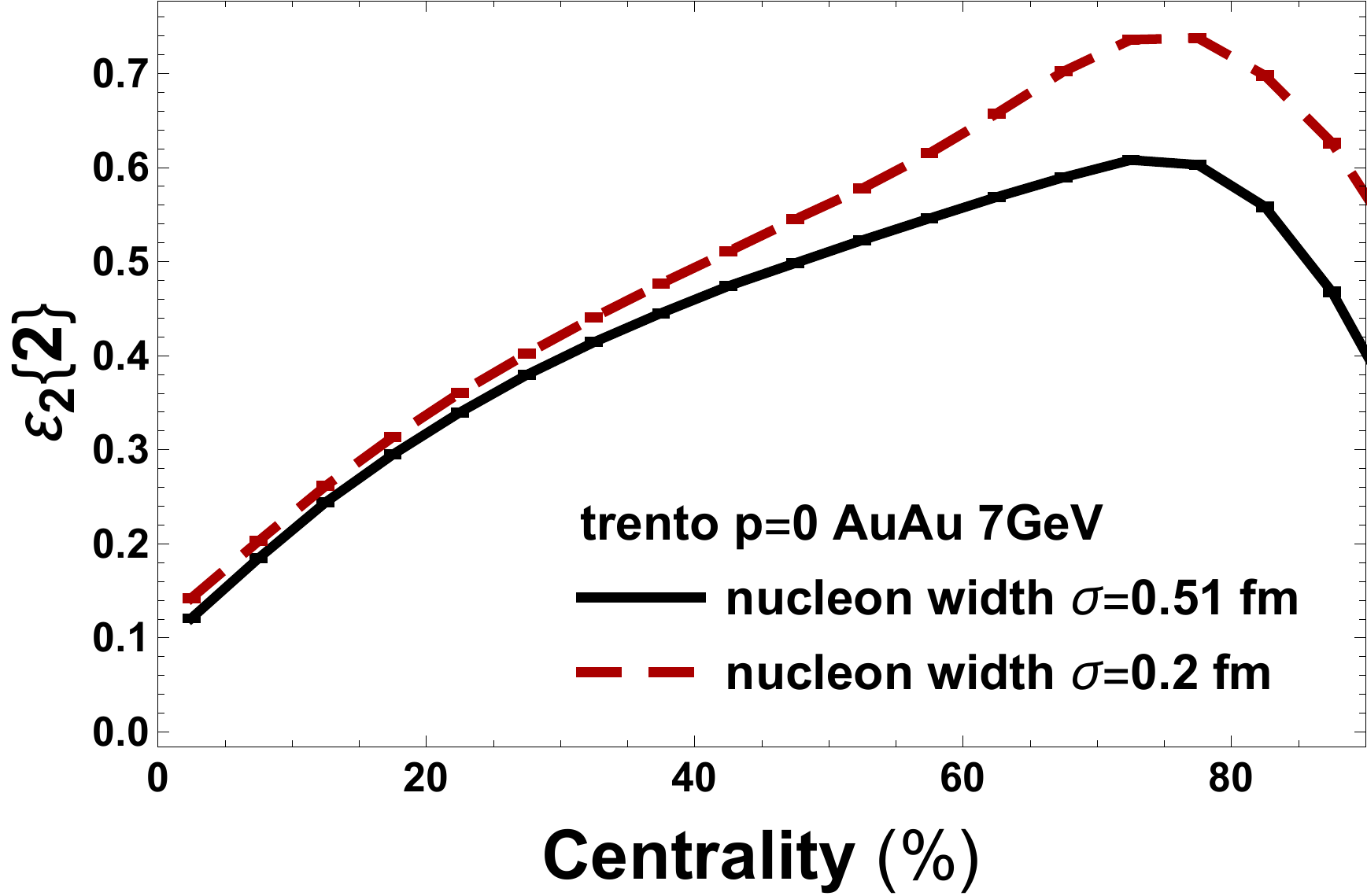} 
	\end{tabular}
	\caption{(Color online) Comparison of $\varepsilon_2\{4\}/\varepsilon_2\{2\}$ (left) and $\varepsilon_\{2\}$ (right) for varied nucleon width, $\sigma$, in Trento where $\sigma=0.51$ fm is the standard value from Bayesian analysis.  Calculations done at AuAu  $\sqrt{s_{NN}}=7$ GeV.}
	\label{fig:sigmabeam}
\end{figure}
%

Finally, we study the influence of the nucleon width parameter $\sigma$ on $\varepsilon_2\{4\}/\varepsilon_2\{2\}$ in Fig.\ \ref{fig:sigmabeam}.  Generally, one would expect that QCD radiation induced by increasing the energy would lead to a slow growth of the nucleon width \cite{Boer:2011fh, Accardi:2012qut}.  Conversely, there could potentially be a small decrease in nucleon width at lower beam energies, although it is unclear {\it{a priori}} how strong this effect might be.  In Fig.\ \ref{fig:sigmabeam}, we find that varying $\sigma$ does have some effect on $\varepsilon_2\{4\}/\varepsilon_2\{2\}$.  From Bayesian analysis in large systems, it was found that a value of $\sigma=0.51~\mathrm{fm}$ for $p=0$ works well in Trento \cite{Bernhard:2016tnd}.  However, in small systems there are some indications that a smaller value of $\sigma=0.3~\mathrm{fm}$ may be preferred \cite{Giacalone:2017uqx}. Thus, if there was a significant change in the size of the nucleons with beam energies, one could expect this to influence $v_2\{4\}/v_2\{2\}$.

%
\section{Beam energy dependence of Linear+cubic response}
\label{sec:kappas}
%

Our next step is to extract the estimator parameters $\kappa_{1,n}, \kappa_{2,n}$ and residuals $\delta_{n}, \Delta_{n,4}$, and to study their beam energy dependence. In Ref.~\cite{sievert:2019zjr} the system size dependence of $\kappa_{1,n}, \kappa_{2,n}$ was extracted, and it was found that for small systems a different type of estimator was needed beyond linear+cubic response.  However, since here we are only considering AuAu collisions we can capture most of the needed physics with linear+cubic response.  Our ignorance of a more complete mapping between initial and final states is quantified in the beam energy dependence of the residuals $\delta_{n}, \Delta_{n,4}$. Finally, we compare the response coefficients extracted from both Trento+v-UPShydro (by which we mean the IP-Glasma-like $p=0$) to actual MC-KLN+v-USPhydro, but we only plot the latter out to $60\%$ centrality due to the lack of statistics there.

%
\begin{figure}[h]
\centering
\begin{tabular}{c c}
\includegraphics[trim=0 150 0 150, clip,width=0.5\linewidth]{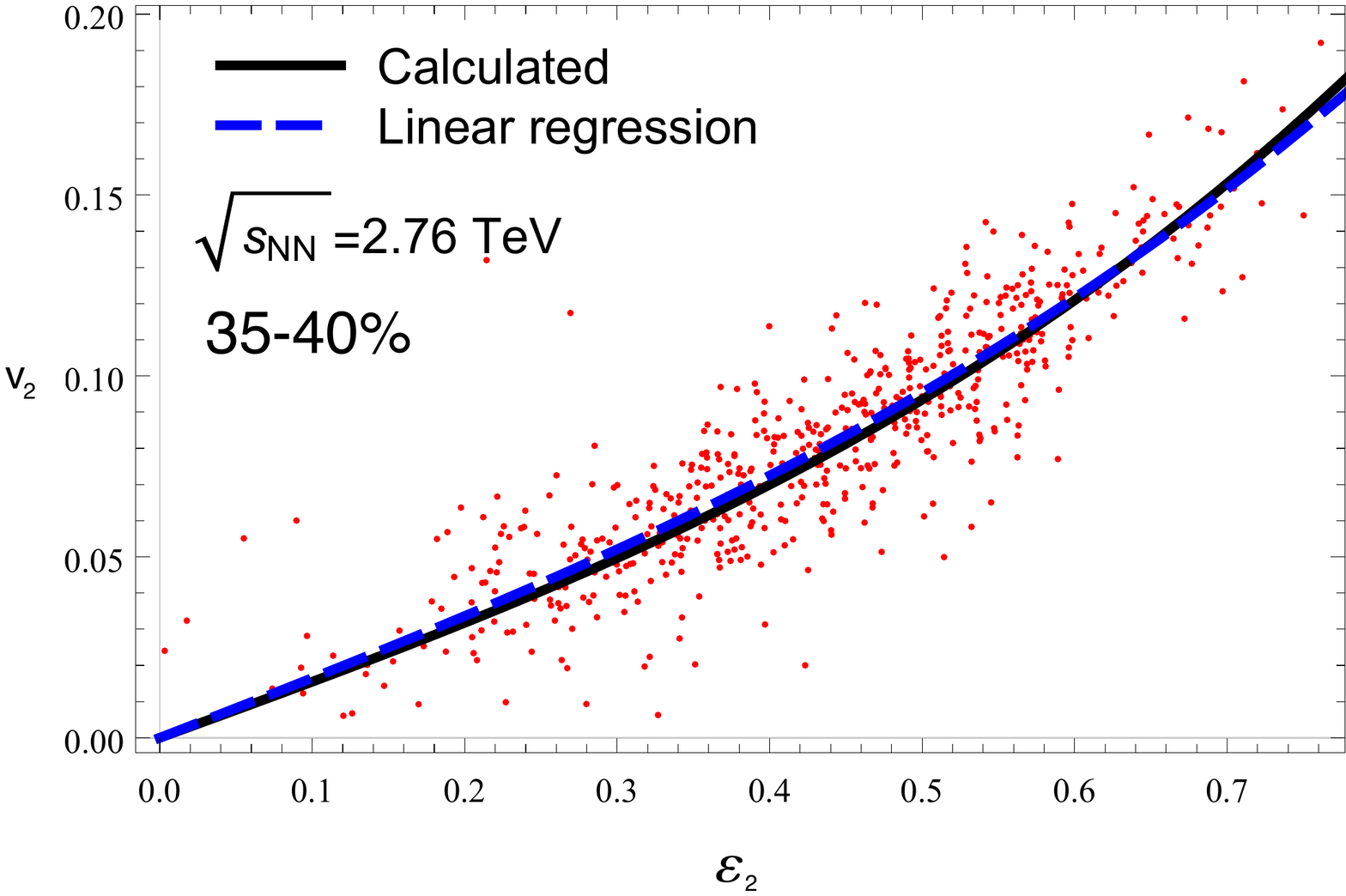} & \includegraphics[trim=0 150 0 150, clip,width=0.5\linewidth]{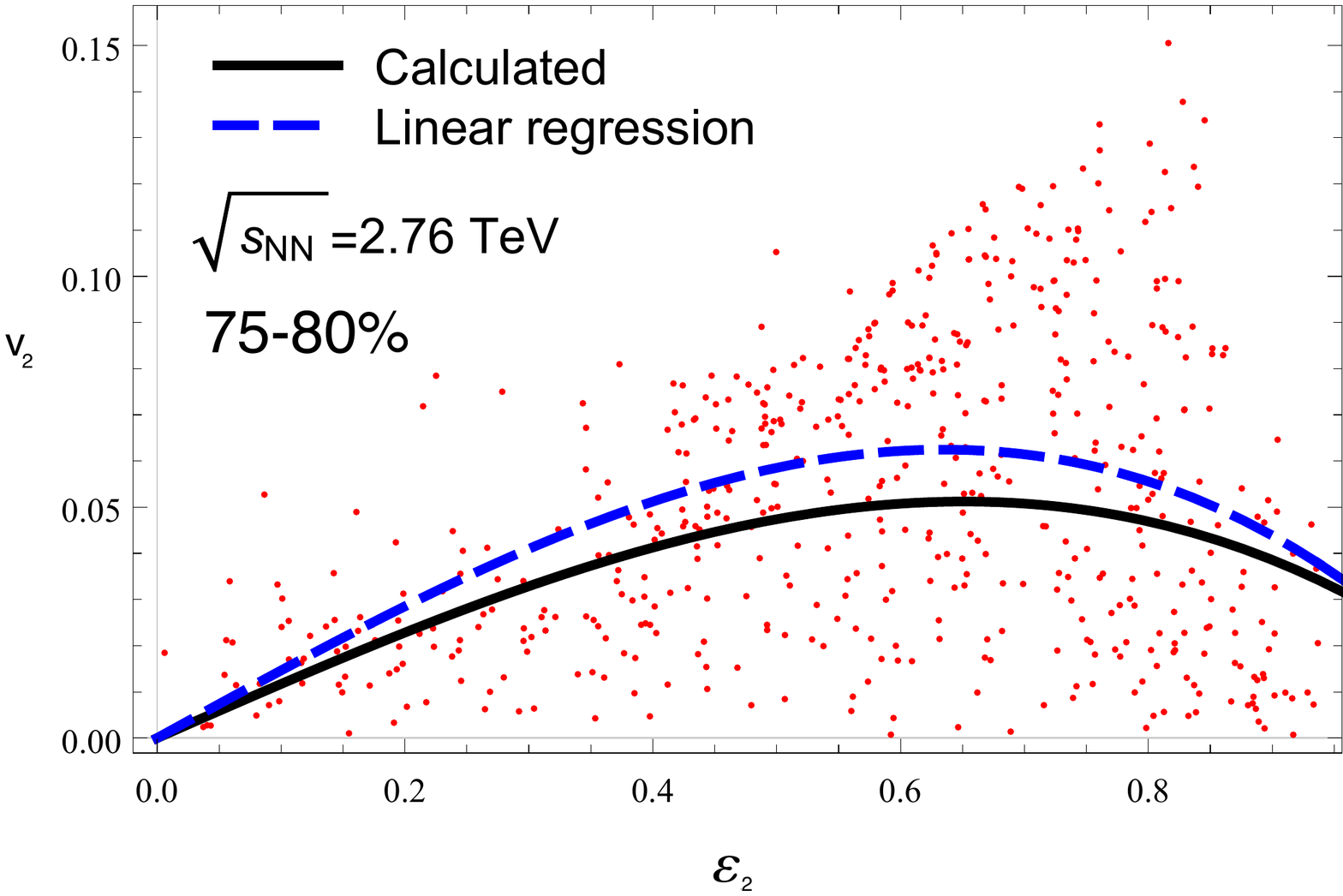} 
\end{tabular}
\caption{(Color online) Comparison of calculated $\kappa_{1,2}$ and $\kappa_{2,2}$ from Eq.\ (\ref{eqn:nonlinear}) to extracted ones using numerical methods shown in mid-central collisions (left) and peripheral collisions (right) shown for Trento+v-USPhydro at PbPb 2.76TeV.}
\label{fig:scatter}
\end{figure}
%

Throughout most of this paper we calculate the estimator coefficients $\kappa_{1,2}$ and $\kappa_{2,2}$ using the expressions derived in Eq.\ (\ref{eqn:nonlinear}), which optimize the linear + cubic estimation for the {\it{vector}} eccentricities $\bm{\mathcal{E}_2}$ and flow harmonics $\bm{V_2}$.  However, it is also interesting to analyze the role played by differences in the $\bm{\mathcal{E}_n}$ and $\bm{V_n}$ event plane angles by comparing the calculation of these coefficients from Eq.\ (\ref{eqn:nonlinear}) versus a fit to only the {\it{magnitudes}} $\varepsilon_2$ and $v_2$.  In Fig.\ \ref{fig:scatter} we see that the two methods are nearly identical for $35-40\%$ mid-central collisions, whereas for $75-80\%$ peripheral collisions, the regression method (magnitudes only) overpredicts the flow $v_2$ arising from the initial $\varepsilon_2$ by about $\sim 20\%$.  Because the calculated method explicitly minimizes the magnitude of the residuals $\langle \delta_n^2 \rangle$, the regression method necessarily provides a less accurate prediction for the final-state flow.

%
\begin{figure}[h]
	\centering
	\begin{tabular}{c c}
		\includegraphics[width=0.5\linewidth]{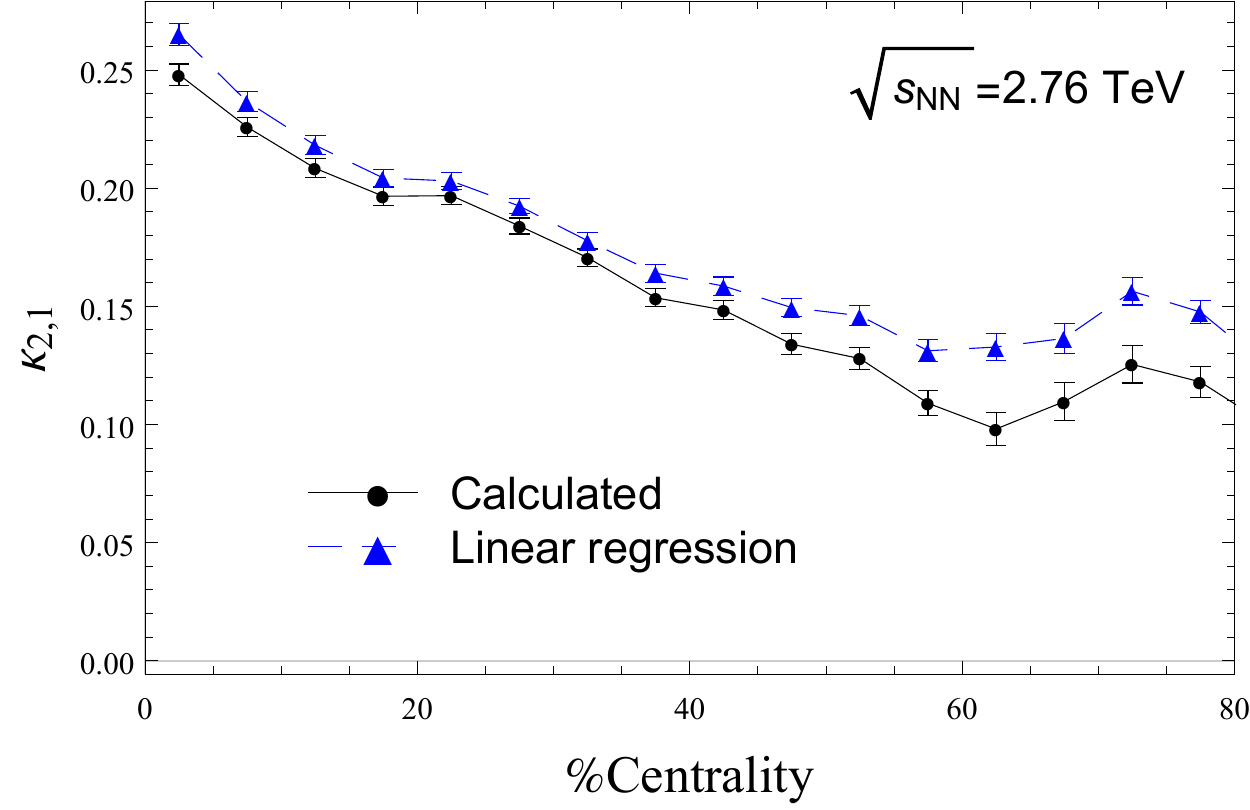} & \includegraphics[width=0.5\linewidth]{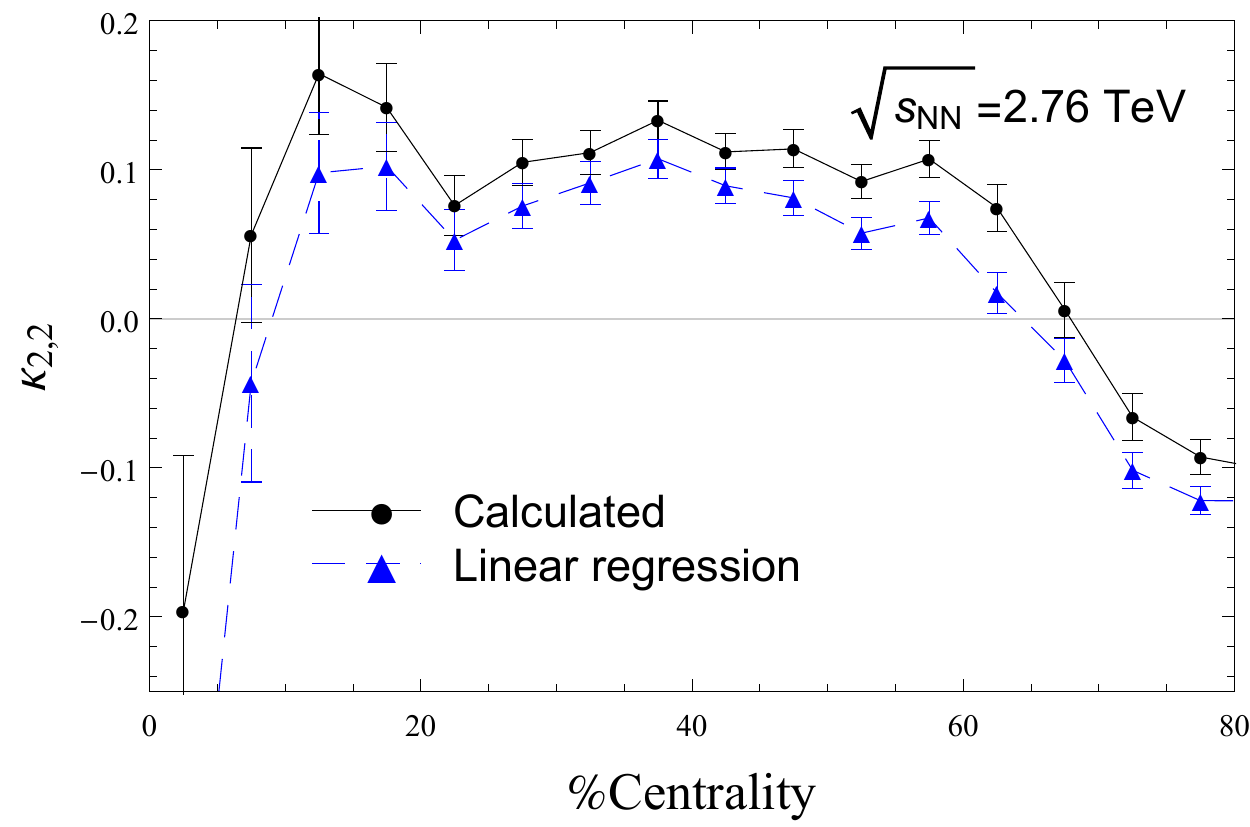} 
	\end{tabular}
	\caption{(Color online) Comparison of calculated $\kappa_{1,2}$ (left) and $\kappa_{2,2}$ (right) from Eq.\ (\ref{eqn:nonlinear}) to extracted ones using numerical methods (linear regression) across centralities shown for Trento+v-USPhydro at PbPb 2.76TeV.}
	\label{fig:linreg}
\end{figure}
%

In Fig.\ \ref{fig:linreg} we show the estimator coefficients $\kappa_{1,2}$ and $\kappa_{2,2}$ extracted from the calculated values \eqref{eqn:nonlinear} and from the magnitude-only regression method.  The trends of the two coefficients are similar for the two methods, with the regression method slightly overestimating the linear contribution and underestimating the cubic contribution.  The over-estimation of the linear coefficient is greatest in peripheral collisions, resulting in the overprediction of the elliptic flow consistent with Fig.\ \ref{fig:scatter}.  From this, we conclude that while comparing the magnitudes of the eccentricities and flow harmonics tells the correct qualitative story about mapping the initial state onto the final state, the event plane angles play an important role in quantitatively constraining these predictors.  For this reason, we will consider only the optimized calculation \eqref{eqn:nonlinear} which includes the event plane angles in forming the linear + cubic map.

\begin{figure}[h]
	\centering
	\begin{tabular}{c c}
		\includegraphics[width=0.5\linewidth]{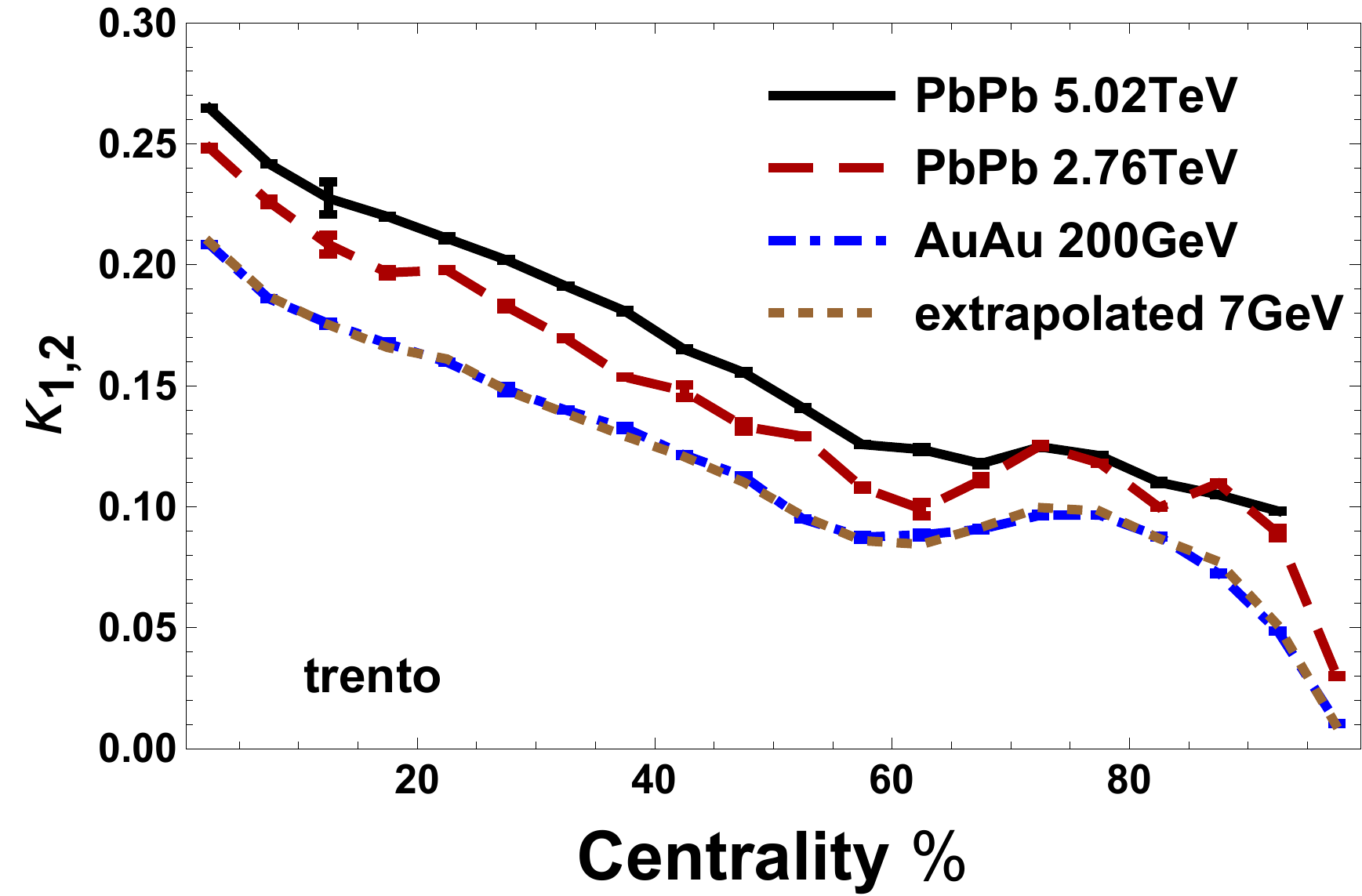} & \includegraphics[width=0.5\linewidth]{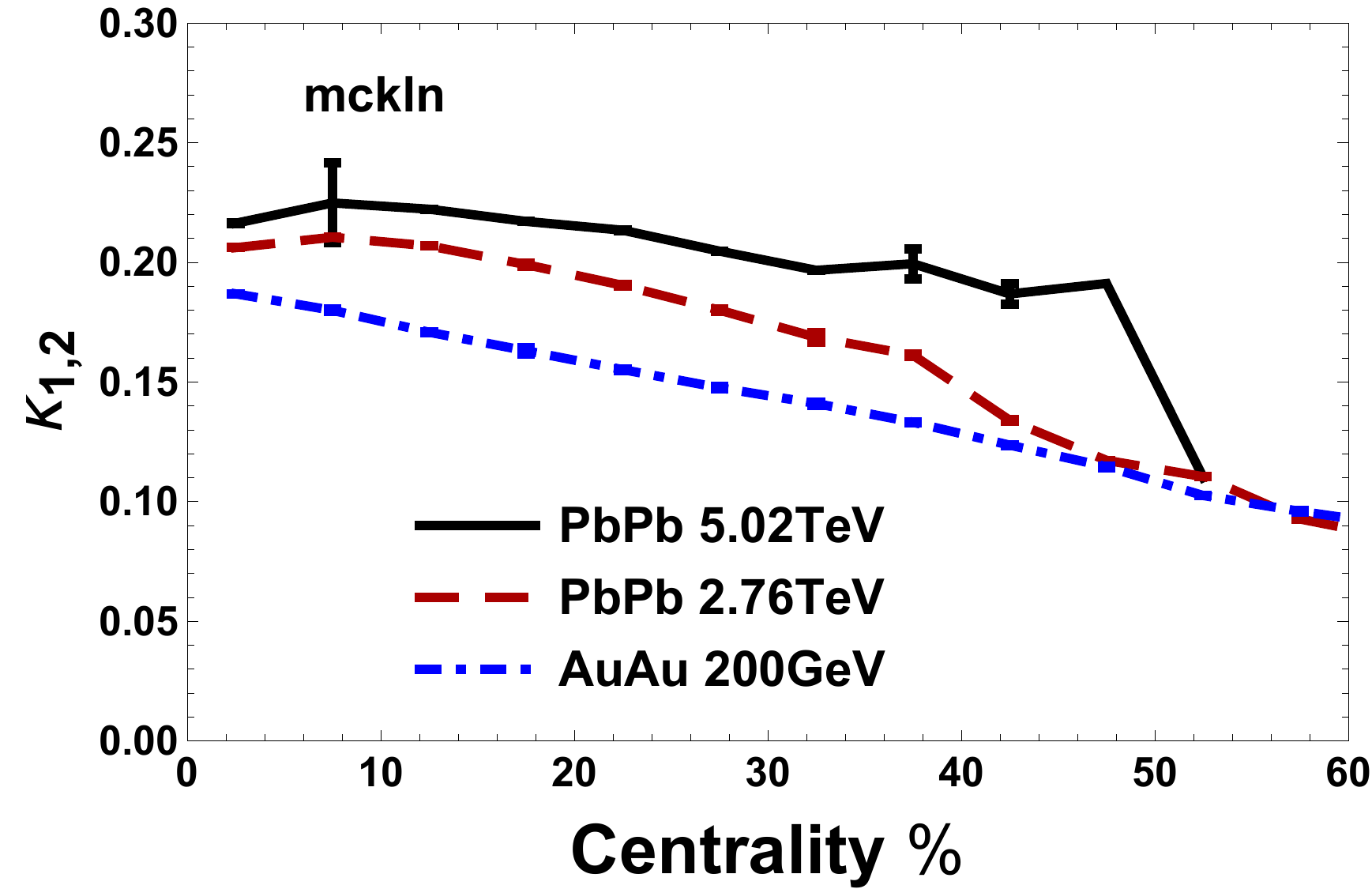} 
	\end{tabular}
	\caption{(Color online) Linear coefficient $\kappa_{1,2}$ for elliptical flow across beam energy extracted using Trento+
		v-USPhydro.  Extrapolated values for AuAu 7 GeV shown in brown dashed lines.
		}
	\label{fig:kappa2}
\end{figure}

\begin{figure}[h]
	\centering
	\begin{tabular}{c c}
		\includegraphics[width=0.5\linewidth]{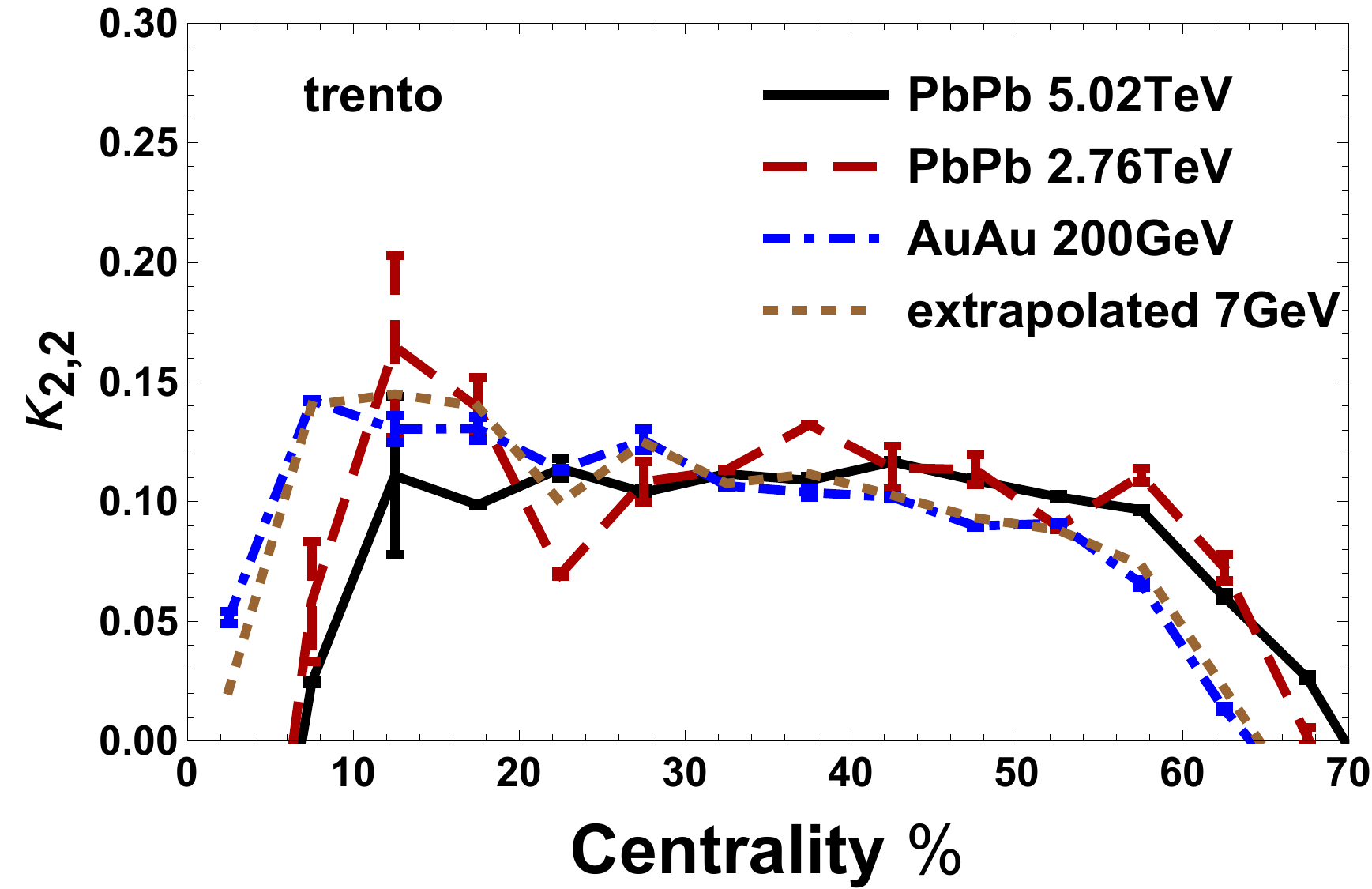} & \includegraphics[width=0.5\linewidth]{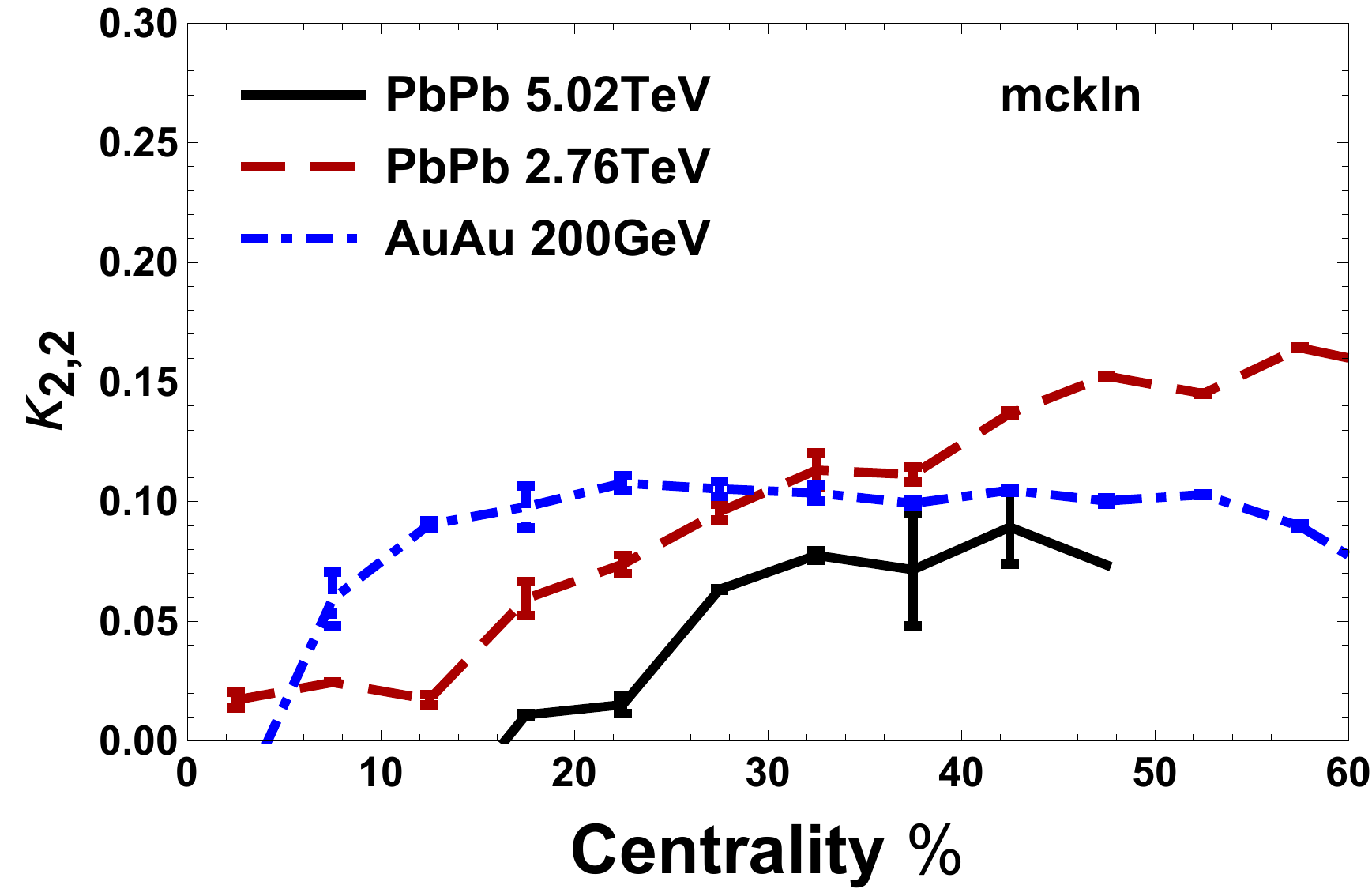} 
	\end{tabular}
	\caption{(Color online) Cubic coefficient $\kappa_{2,2}$ for elliptical flow across beam energy extracted using Trento+
		v-USPhydro.  Extrapolated values for AuAu 7 GeV shown in brown dashed lines
	}
	\label{fig:kappa22}
\end{figure}

\begin{figure}[h]
	\centering
	\begin{tabular}{c c}
		\includegraphics[width=0.5\linewidth]{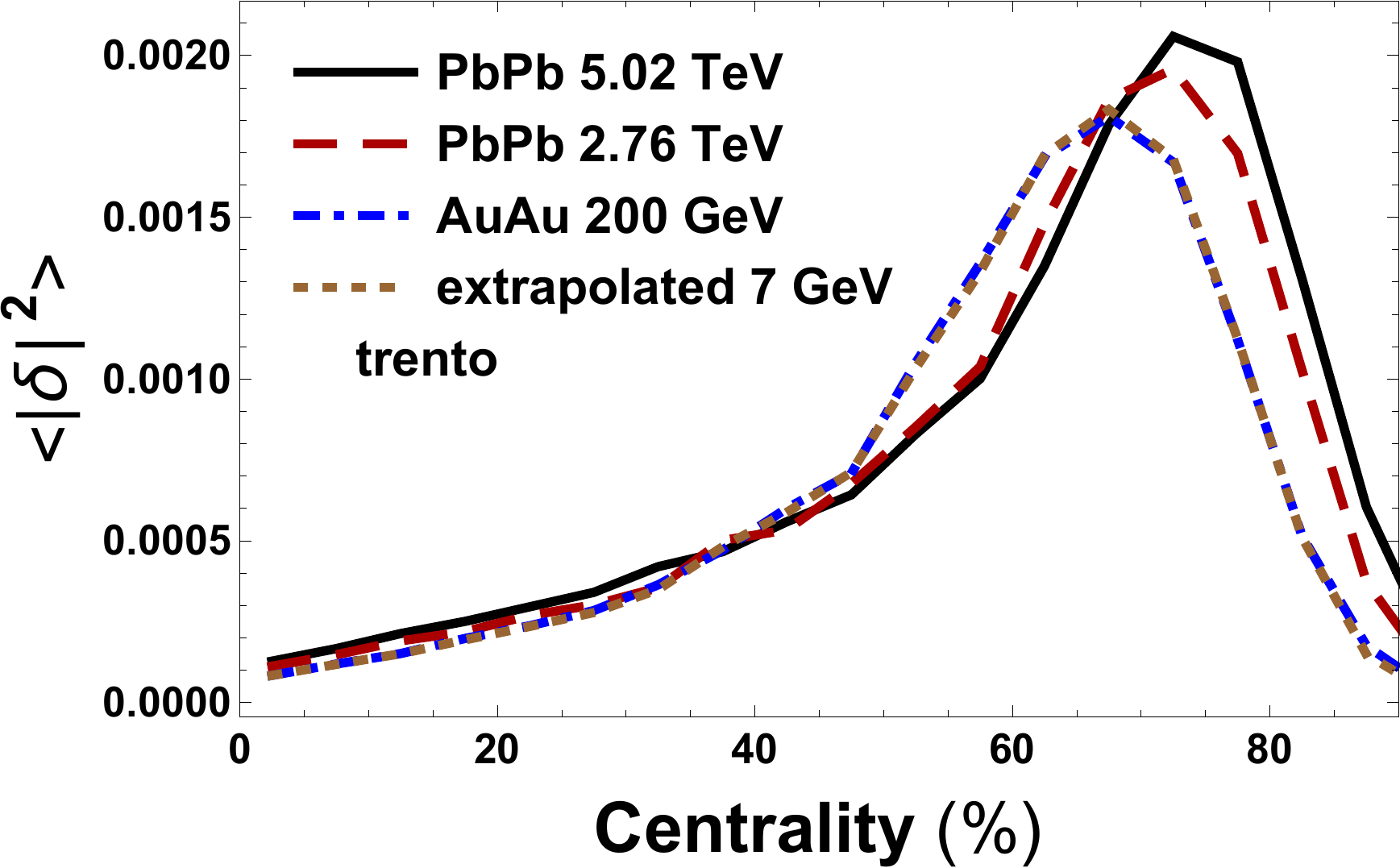} & \includegraphics[width=0.5\linewidth]{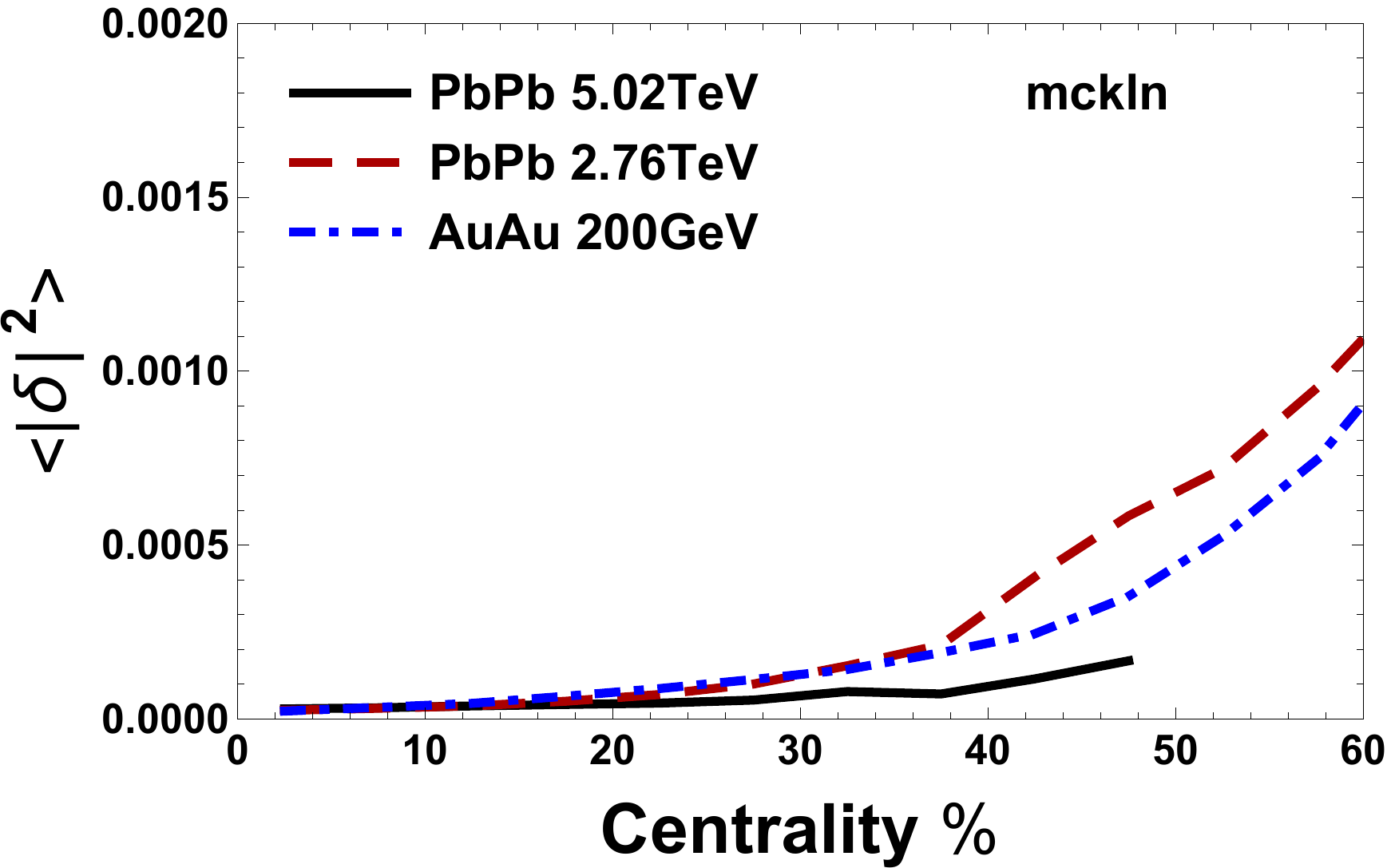} 
	\end{tabular}
	\caption{(Color online) Residual contribution to $\langle v_2^2\rangle$ for elliptical flow across beam energy extracted using Trento+
		v-USPhydro.  Extrapolated values for AuAu 7 GeV shown in brown dashed lines.
	}
	\label{fig:delta2}
\end{figure}

\begin{figure}[h]
	\centering
	\begin{tabular}{c c}
		\includegraphics[width=0.5\linewidth]{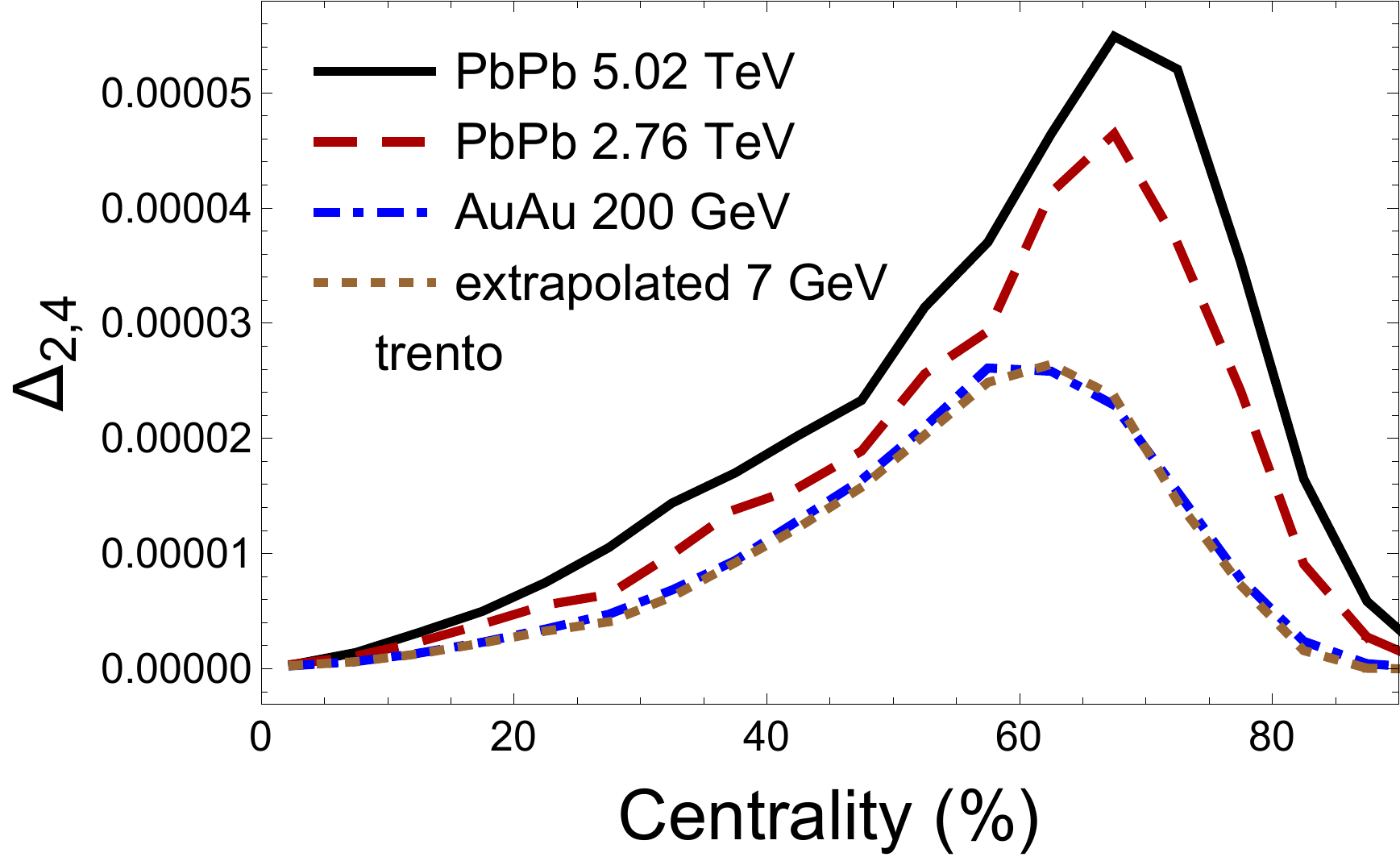} & \includegraphics[width=0.5\linewidth]{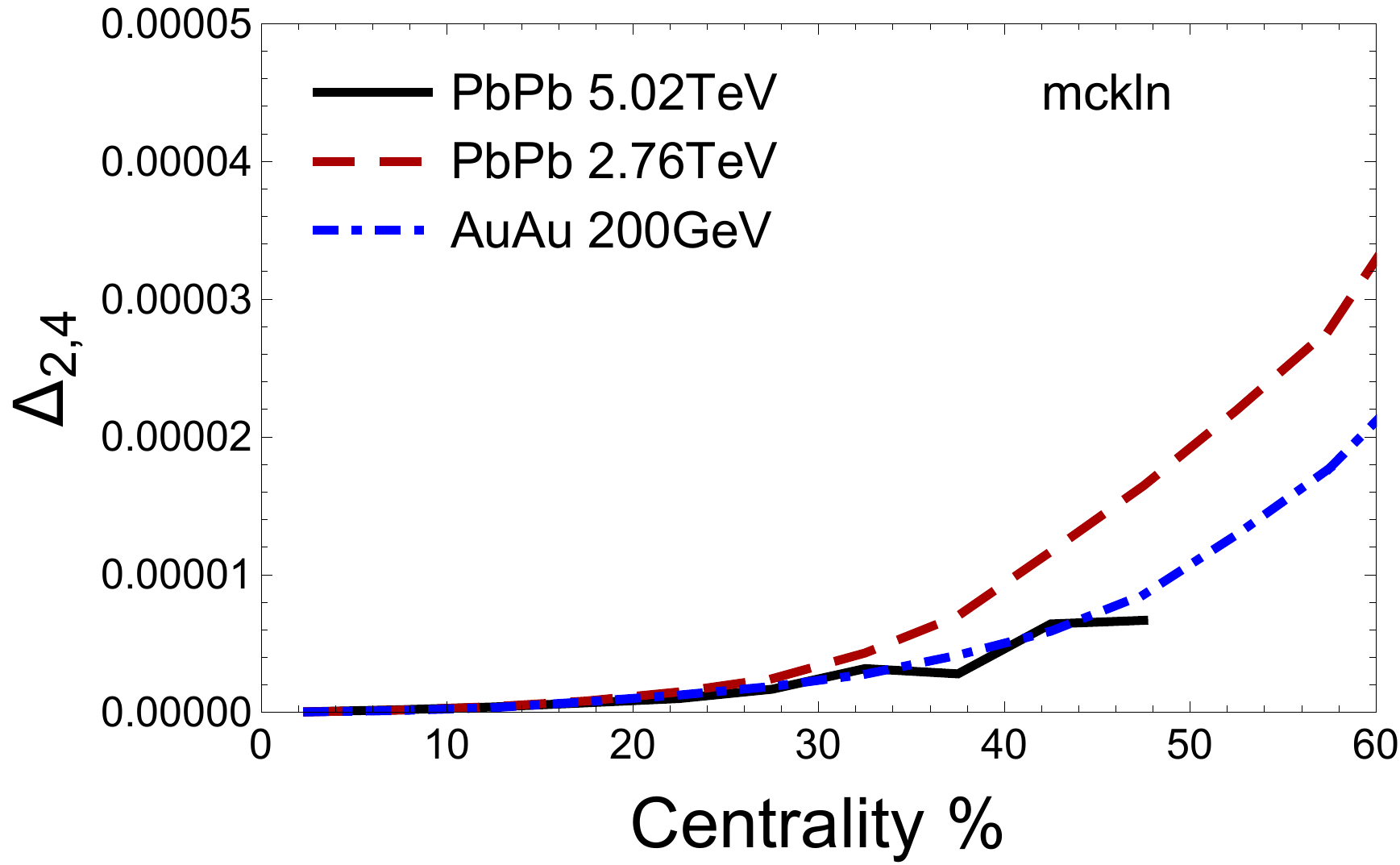} 
	\end{tabular}
	\caption{(Color online) Residual contribution to $\langle v_2^4\rangle$ for elliptical flow across beam energy extracted using Trento and MC-KLN +
		v-USPhydro.  Extrapolated values for AuAu 7 GeV shown in brown dashed lines
	}
	\label{fig:delta4}
\end{figure}

\begin{figure}[h]
	\centering
	\begin{tabular}{c c}
		\includegraphics[width=0.5\linewidth]{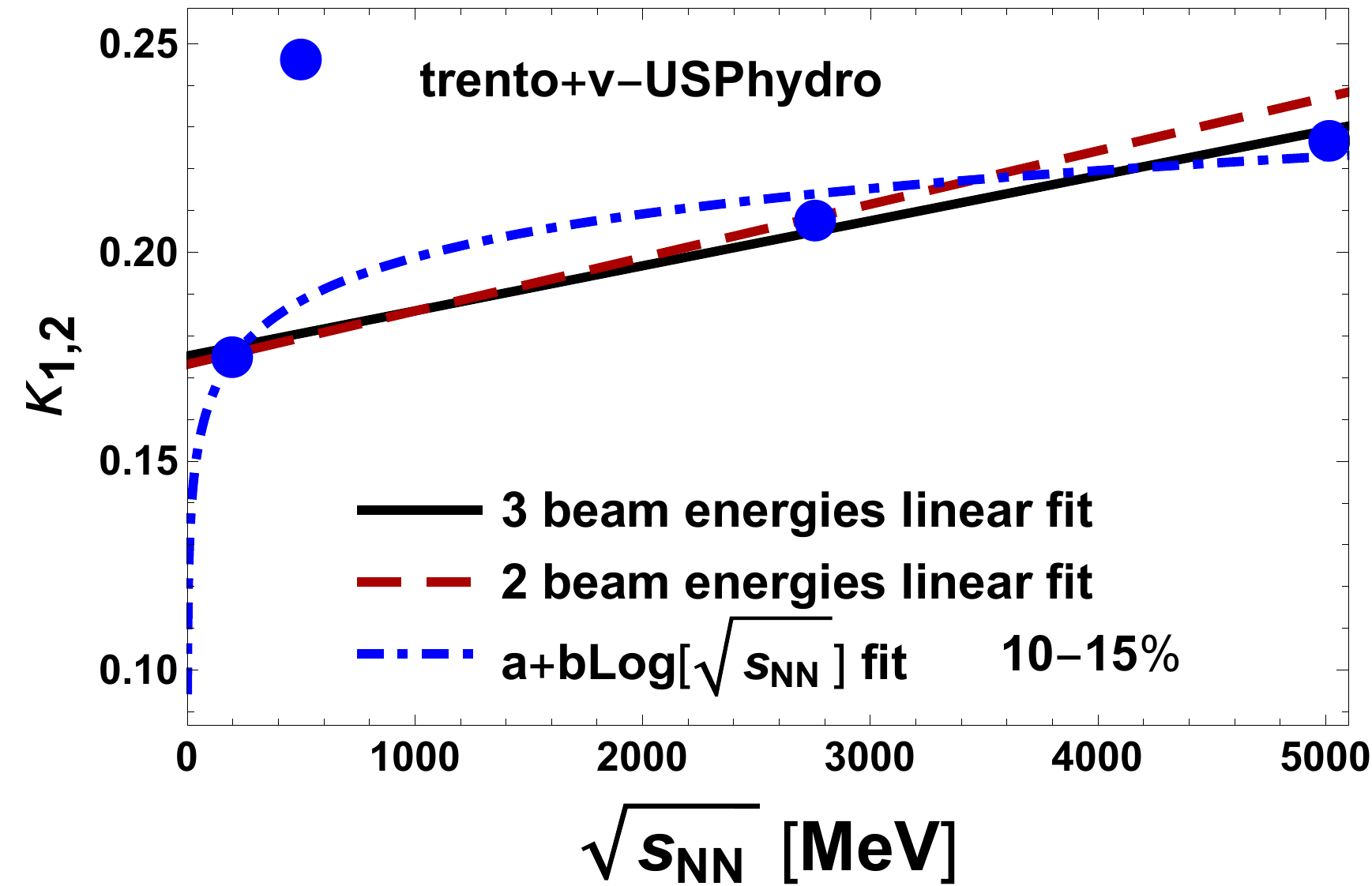} 
	\end{tabular}
	\caption{(Color online) Extrapolation of $\kappa_{1,2}$ down to lower beam energies using a linear fit either considering all 3 beam energies, linear fit with  just the lowest 2, or a fit with the format $a+b\log[\sqrt{s_{NN}}]$ where $a$ and $b$ are constants.  The calculations here use initial conditions generated by Trento with $p=0$, and the plot is for the $10-15\%$ centrality bin.}
	\label{fig:extrap}
\end{figure}

In Figs.\ \ref{fig:kappa2}-\ref{fig:delta4} we plot the estimator parameters and residuals for the top-energy heavy ion collisions at RHIC and the LHC (AuAu $200~\mathrm{GeV}$, PbPb $2.76~\mathrm{TeV}$, and PbPb $5.02~\mathrm{TeV}$).  We also include these quantities linearly extrapolated down to AuAu $7~\mathrm{GeV}$ using the approaches as described in Sec.\ \ref{sec:extract} and as visualized in Fig.~\ref{fig:extrap}.  In Sec.\ \ref{sec:extract} we compare a few different approaches to extrapolating these coefficients down to lower beam energies, but let us note for now that the different methods appear to be relatively robust, regardless of the extrapolation method.   

%
%

In Fig.\ \ref{fig:kappa2} the calculated linear response coefficients are shown for both the Trento $p=0$ and MC-KLN models.  While there are some differences between the two models, the orders of magnitude are comparable, and both illustrate that $\kappa_{1,2}$ is anticorrelated with centrality class.  Additionally, both models exhibit a hierarchy with higher beam energies generating larger linear response coefficients.  This hierarchy is clearest in central collisions but shows some signs of converging in peripheral collisions.  This strong energy hierarchy indicates that decreasing the beam energy suppresses linear response. 

%
%

Similarly, in Fig.\ \ref{fig:kappa22} we show the cubic response coefficients $\kappa_{2,2}$.  For Trento, $\kappa_{2,2}$ appears to see little to no change across beam energy.  We reiterate that in Trento, hydrodynamics was able to reproduce the experimental data with a single fixed set of parameters such as the shear viscosity $\eta/s\sim 0.05$.  In contrast, for  MC-KLN it was necessary to vary the shear viscosity $\eta/s$ across beam energy in order to reproduce experimental data; as a result, the cubic response coefficient exhibits a much stronger beam energy dependence.  This suggests that the cubic response coefficients are sensitive to energy-dependent changes in the medium parameters.

%
%

Unlike the estimator coefficients $\kappa_{1,n} , \kappa_{2,n}$ which have definite interpretations as originating from the properties of the final state, the residuals
$\langle \delta_n^2\rangle$ as defined in Eq.~\eqref{e:resdef} are measures of our ignorance.  As was previously found in Ref.~\cite{Giacalone:2017uqx}, we see in Fig.\ \ref{fig:delta2} that the absolute magnitudes of the optimized residuals are rather small, indicating that the linear+cubic estimator is performing well.\footnote{Note however from Eq.~\eqref{e:optcum2} that the contributions of the residuals to the cumulants scale more closely with $\sqrt{\langle \delta_n^2 \rangle}$ and $\sqrt[4]{\Delta_{n,4}}$ than with the residuals themselves, so one should not infer that these corrections are vanishingly small.}  The residuals do, however, grow significantly for peripheral collisions; this is consistent with the picture from the deterioration of the Pearson coefficients $Q_n$ from Eq.\ \eqref{eqn:pear} in that region.  These trends are comparable between Trento and MC-KLN.  In both models, the variation in the residuals with beam energy is small, particularly in central collisions.  In peripheral collisions a discernible splitting between the energy dependence of the residuals is seen, but the behavior is quite different for Trento and MC-KLN.  For Trento there is a systematic hierarchy with the residuals at lower beam energies being somewhat smaller than at higher energies.  In contrast, the residuals for MC-KLN are not ordered with beam energy; they do, however, track the different values of $\eta/s$.  For instance, PbPb $2.76~\mathrm{TeV}$ has the largest viscosity of $0.11$ and also has the largest residual, whereas PbPb $5.02~\mathrm{TeV}$ has the smallest viscosity of $0.05$ and also a very small residual.

%

In Fig.\ \ref{fig:delta4} the new residual $\Delta_{2,4}$ contribution is shown, which has a similar centrality dependence to what is seen in Fig.\ \ref{fig:delta2} for $\langle \delta_2^2\rangle$.  For $p=0$, both residuals peak around $60-70\%$ centrality, and lower beam energies correlate with smaller residuals with the peak shifting towards small centrality classes. The energy dependence for MC-KLN is also similar to what was seen in Fig.\ \ref{fig:delta4}, and in all cases the residual $\Delta_{2,4}$ is very small.

%
\begin{figure}[h]
	\centering
	\begin{tabular}{c c}
		\includegraphics[width=0.5\linewidth]{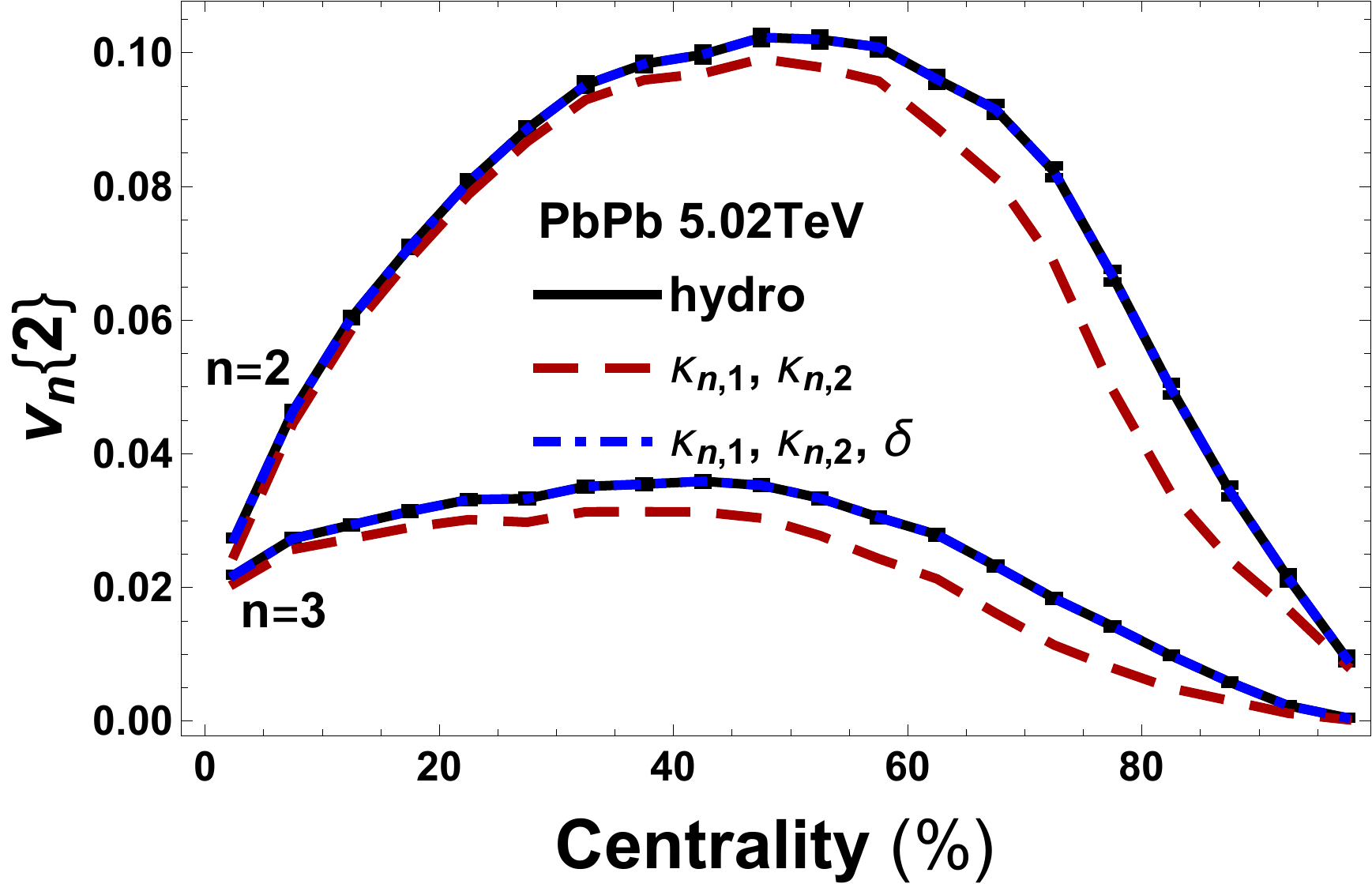} 
	\end{tabular}
	\caption{(Color online) Direct Trento+v-USPhydro hydrodynamical calculations (solid black) versus the predicted $v_n\{2\}$ from linear+cubic response (red long dashed) and the reconstructed $v_n\{2\}$ from linear+cubic response+residual (blue short dashed). Calculations are performed for PbPb $5.02~\mathrm{TeV}$ with initial conditions using Trento $p=0$.
	}
	\label{fig:recon}
\end{figure}
%

Having determined the response coefficients, we can visualize directly how well the linear+cubic estimator is able to reproduce the final flow harmonics across beam energies.  In Fig.\ \ref{fig:recon} we compare the linear+cubic estimator Eq.\ \eqref{eqn:precubic} from the initial state with the calculated flow harmonics, and by adding the residuals back in, we can fully account for the final state flow.  As seen previously, we again observe that the linear+cubic estimator is least successful in peripheral collisions, where the residual makes a more significant contribution to the final state flow harmonics.

%
\begin{figure}[h]
	\centering
	\includegraphics[width=0.5\linewidth]{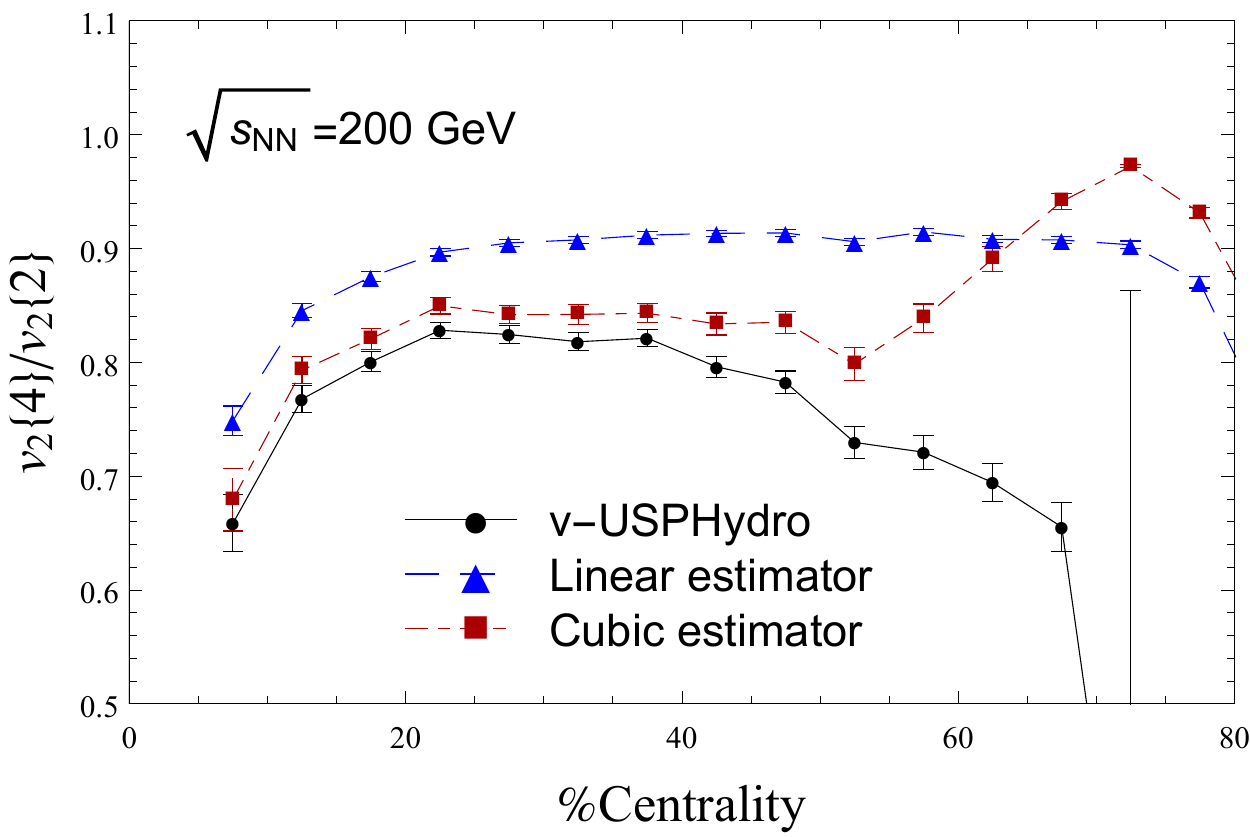} \\
	\includegraphics[width=0.5\linewidth]{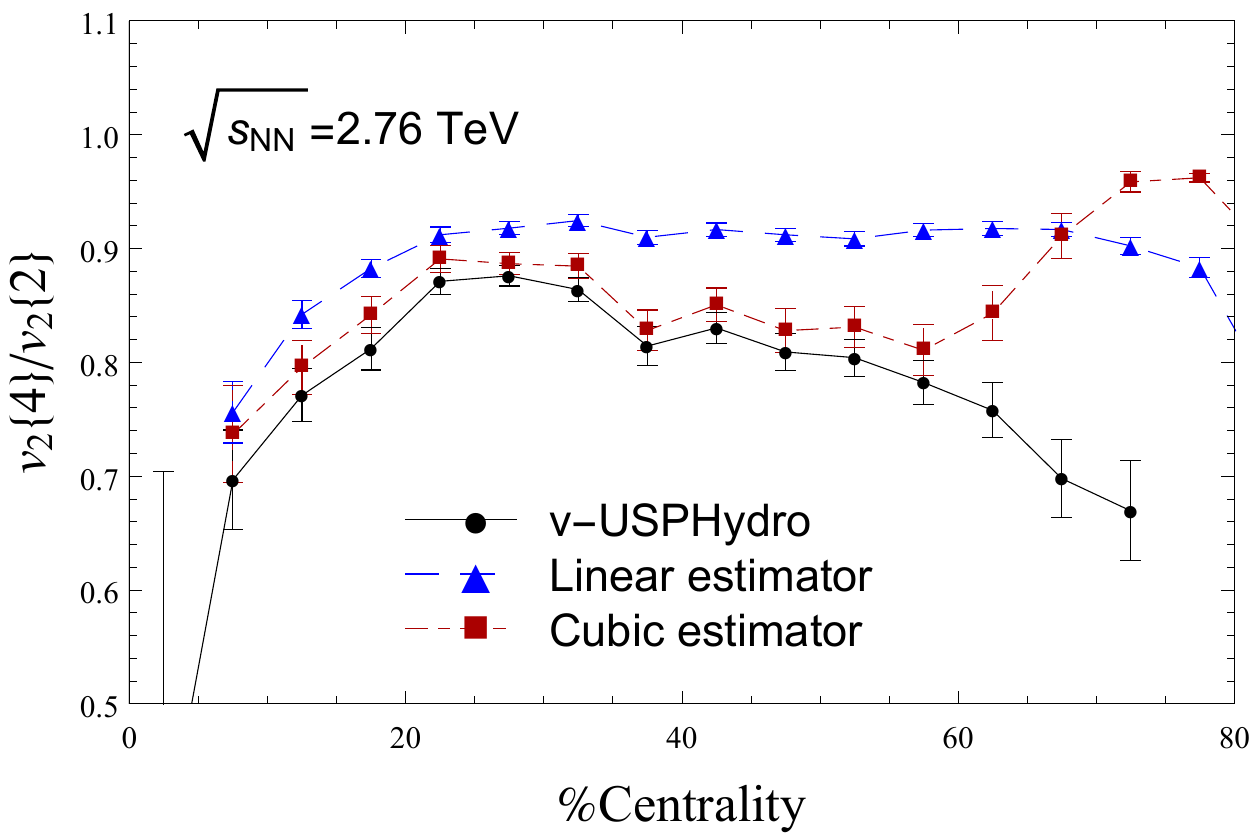} \\
	\includegraphics[width=0.5\linewidth]{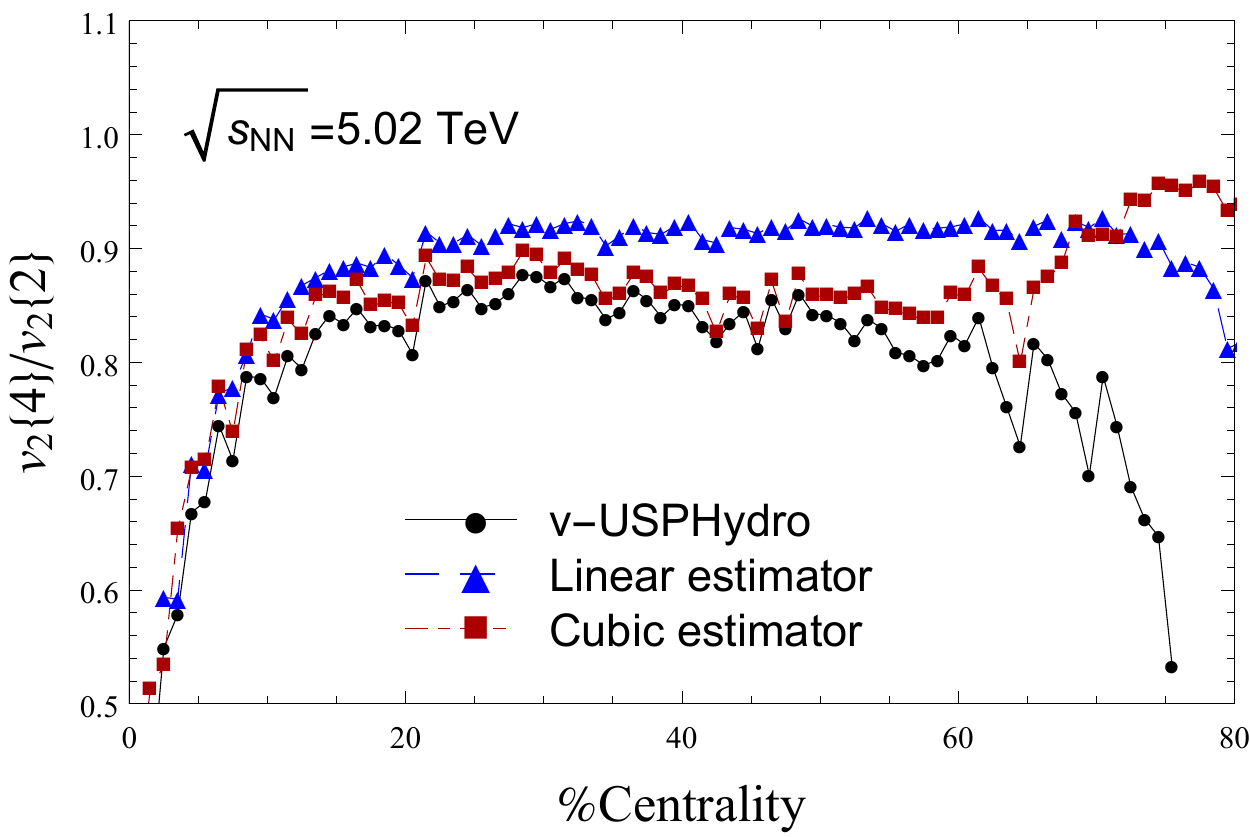} 
	\caption{The cumulant ratio $v_2 \{4\} / v_2 \{2\}$ calculated for the top RHIC and LHC energies using full hydrodynamics (black), linear response only (blue), and linear+cubic response (red).  Calculations are performed for AuAu $200~\mathrm{GeV}$ collisions (top), PbPb $2.76~\mathrm{TeV}$ collisions (middle), and PbPb $5.02~\mathrm{TeV}$ collisions (bottom) with initial conditions using Trento $p=0$.}
	\label{fig:lincubv2fluc}
\end{figure}
%

Similarly, in Fig.\ \ref{fig:lincubv2fluc} we plot the cumulant ratio $v_2 \{4\} / v_2 \{2\}$ which is a measure of the event-by-event fluctuations of the elliptic flow harmonic $v_2$ as shown in Eq.~\eqref{e:cumratio}.  Here we compare the fluctuations as predicted from just a pure linear estimator ($\kappa_{2,2} = 0$) with those using a linear+cubic estimator \eqref{eqn:precubic} and the full final-state flow after running hydrodynamics.\footnote{Note that the optimized coefficients for linear response {\it{only}} differ from the expressions for $\kappa_{1,n}$ given in Eq.~\eqref{eqn:nonlinear}.  Here we have used the correct optimized linear response coefficients as derived in Ref.~\cite{Noronha-Hostler:2015dbi}.}  For all three cases at top RHIC and LHC energies, the linear+cubic response reproduces well the flow fluctuations from $0-60\%$ centralities.  From these plots we conclude that the contributions from the residuals is of $\ord{5\%}$ and likely within the uncertainties of the measurement.  As such, when we later make predictions for $v_2\{4\}/v_2\{2\}$ at lower beam energies we will not include the residuals.

%
%

%
\section{Extrapolation to low beam energies}
\label{sec:extract}
%

The last step is to use the information we have computed above for the top three RHIC and LHC energies in order to extrapolate to lower beam energies.  Knowing that all of these model calculations incorporate the weak energy dependence present at these kinematics, we take a naive approach and simply compute a linear fit to the top beam energies.  This is illustrated for the energy dependence of the linear response coefficient $\kappa_{1,2}$ in Fig.\ \ref{fig:extrap}.  On these scales, the AuAu $200~\mathrm{GeV}$ data from RHIC sit at less than $10\%$ the energy of the LHC data, so a downward linear extrapolation will change the parameters very little from AuAu $200~\mathrm{GeV}$ to $7~\mathrm{GeV}$, as seen in Fig.\ \ref{fig:extrap}.  We compare a few different approaches to such an extrapolation, including a linear fit to all three energies versus to only the lower two energies.  Incorporating the statistical error bars into the extrapolation, as opposed to using only the central values, also did not make a significant difference in the extrapolated parameters.  

Given that we expect new physical mechanisms to play a role in the medium description at the lower beam energies, it seems appropriate to also consider a fit which deviates steeply from linearity at low energies.  As a toy model for the onset of such behavior at low energies, we also compare to a log fit $a+b\log[\sqrt{s_{NN}}]$, where $a$ and $b$ are constants.  It may be that other functions which deviate significantly at low energies would work better, but given the present lack of published data in this region it is impossible to make a more sophisticated fit.  Incorporating new data from additional beam energies will certainly help in constraining these extrapolations, which we leave for future work.

While this naive extrapolation is far from able to encapsulate all the physics required to study the lowest beam energies, there are certain aspects that this approach does capture.  For instance, basing the mapping coefficients on the higher beam energies will ensure that a shorter lifetime at lower beam energies is taken into account.  Since collisions at $5.02~\mathrm{TeV}$ reach significantly higher maximum temperatures ($T\sim 600~\mathrm{MeV}$ vs. $T\sim 400~\mathrm{MeV}$ at $200~\mathrm{GeV}$) and we keep the freeze-out temperature fixed at $T_{FO}=150~\mathrm{MeV}$, extrapolating downwards to lower beam energies ensures that the hydrodynamic lifetime is shorter. Additionally, our results will be applicable in the region of the phase diagram where $\eta/s\sim const$.

%
\section{Results}
\label{sec:results}
%

Using our newly extrapolated transport coefficients, we are now able to make baseline predictions for the lower beam energies.  While preliminary results have been shown in a proceedings from STAR \cite{Magdy:2018itt}, we are still waiting on the final published data to make direct comparisons to our results.  (In fact, as discussed below, we may be able to use the published data to {\it{extract}} these coefficients.)  

%
\begin{figure}[h]
	\centering
	\begin{tabular}{c c}
		\includegraphics[width=0.5\linewidth]{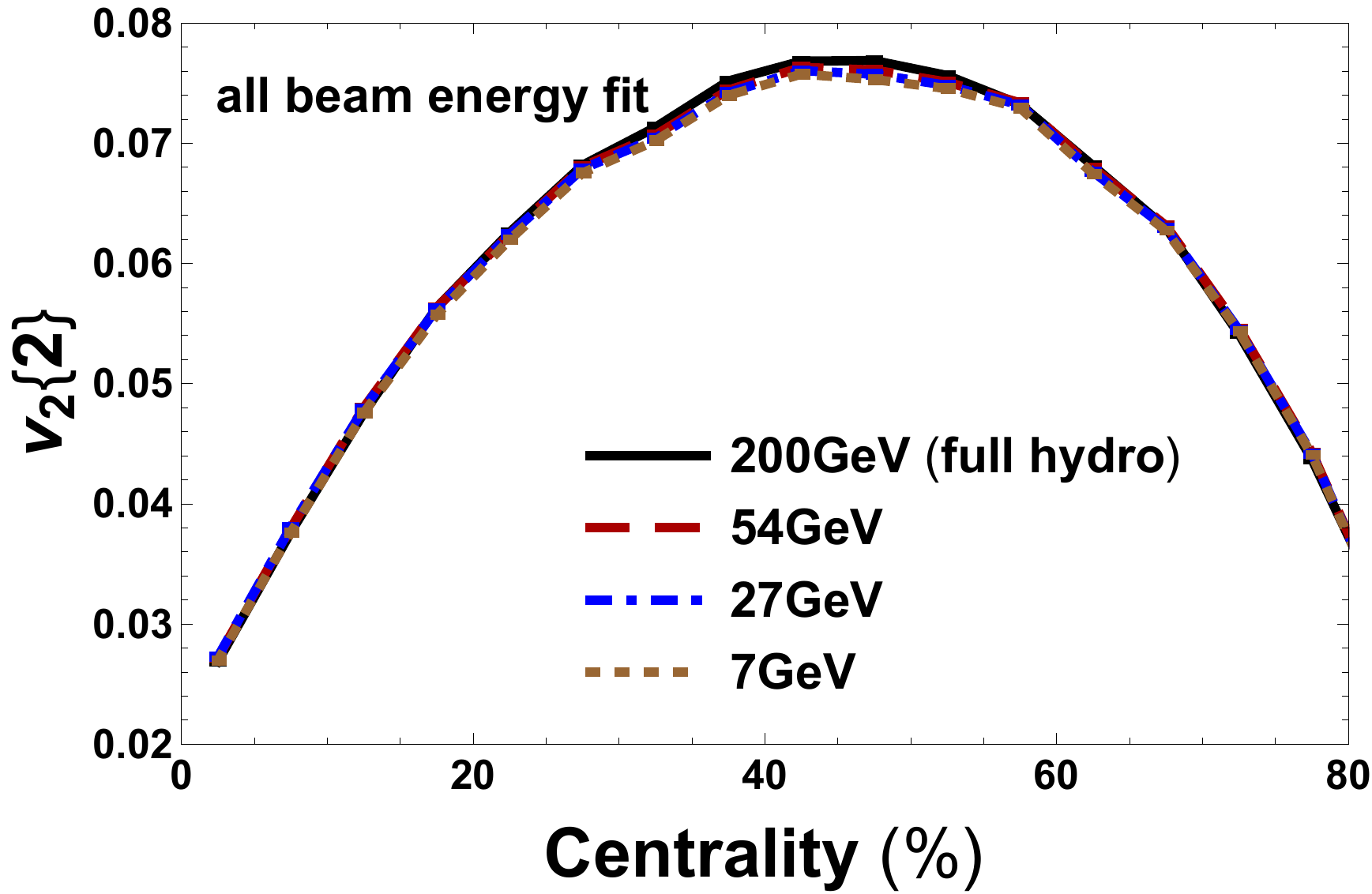} & \includegraphics[width=0.5\linewidth]{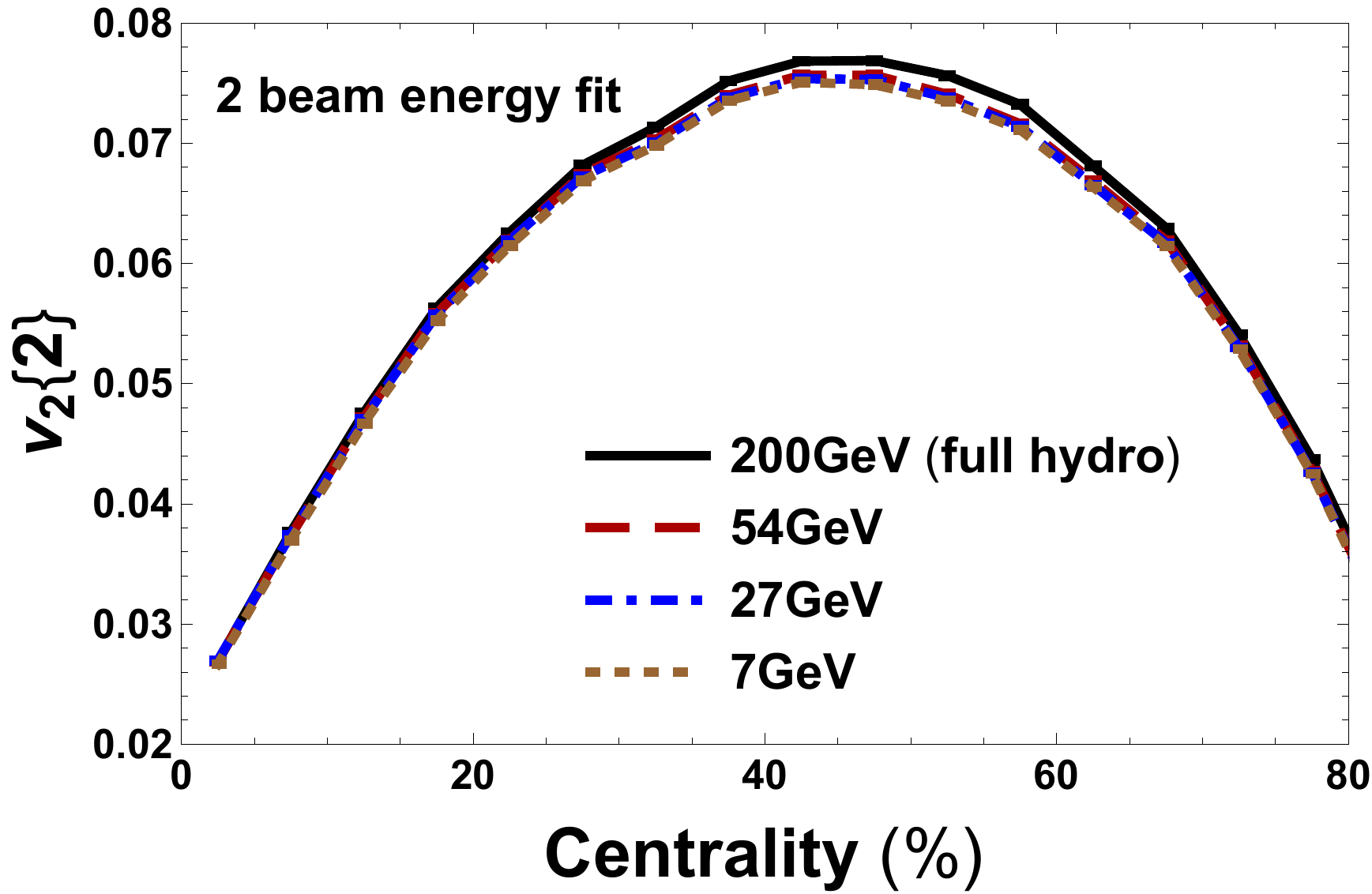} 
	\end{tabular}
	\caption{(Color online) Direct Trento+v-USPhydro hydrodynamic calculations for AuAu $200~\mathrm{GeV}$ (solid black) versus the predicted $v_2\{2\}$ from extrapolating the linear+cubic response coefficients and residual to lower energies.  Here we compare a linear extrapolation to fits of all beam energies (left) versus lowest two beam energies (right).  Baseline predictions are made for $54~\mathrm{GeV}$ (red long dashed), $27~\mathrm{GeV}$ (blue dot dashed), and $7~\mathrm{GeV}$ (brown short dashed).}
	\label{fig:linv22}
\end{figure}
%

We first consider the linear extrapolations as shown in Fig.\ \ref{fig:linv22}, where the fit using all three top RHIC + LHC energies is shown on the left and the fit to only the $200~\mathrm{GeV}$ and $2.76~\mathrm{TeV}$ energies is shown on the right.  We use these fits to extrapolate the two-particle cumulant down to $54~\mathrm{GeV}$, $27~\mathrm{GeV}$, and $7~\mathrm{GeV}$ by using Trento $p=0$ initial conditions at these energies, together with the extrapolated values of $\kappa_{1,2}$, $\kappa_{2,2}$, and $\langle \delta_2^2 \rangle$.  For the extrapolations based on both the all-energy and two-energy fits, there is almost no beam energy dependence.  This is consistent with both the expected weak energy dependence in the eikonal approximation (see Fig.~\ref{f:xsecplot}) to the initial state eccentricities (see Figs.~\ref{fig:eccps}-\ref{fig:sigmabeam}) and to the small lever arm to extrapolate down from $200~\mathrm{GeV}$ (see Figs.~\ref{fig:kappa2}-\ref{fig:delta4}).  We do not show the cumulant ratio $v_2\{4\}/v_2\{2\}$ for these fits because the results look nearly identical across beam energies, regardless of the method of linear extrapolation.

%
\begin{figure}[h]
	\centering
	\begin{tabular}{c c}
		\includegraphics[width=0.5\linewidth]{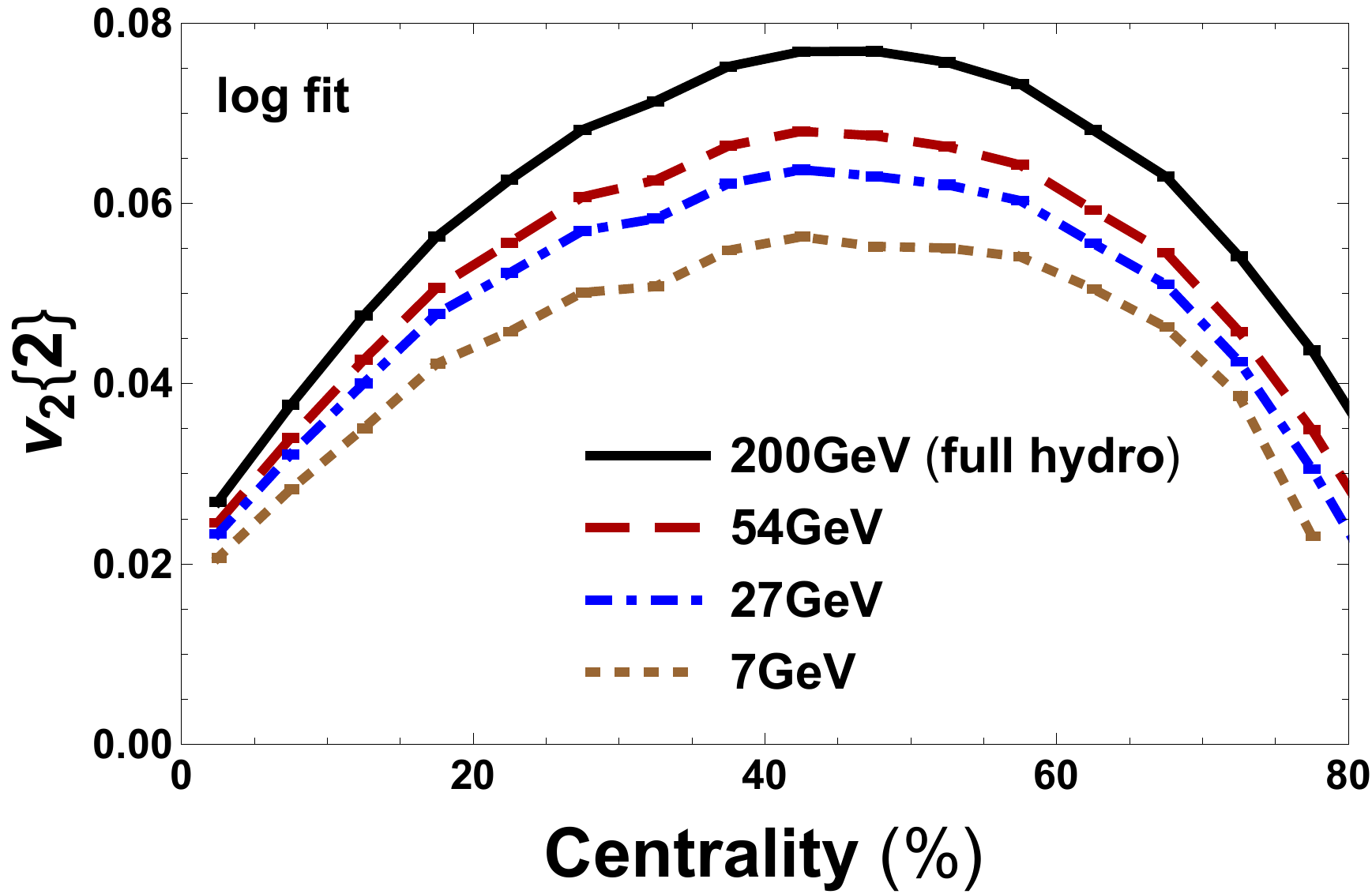} & \includegraphics[width=0.5\linewidth]{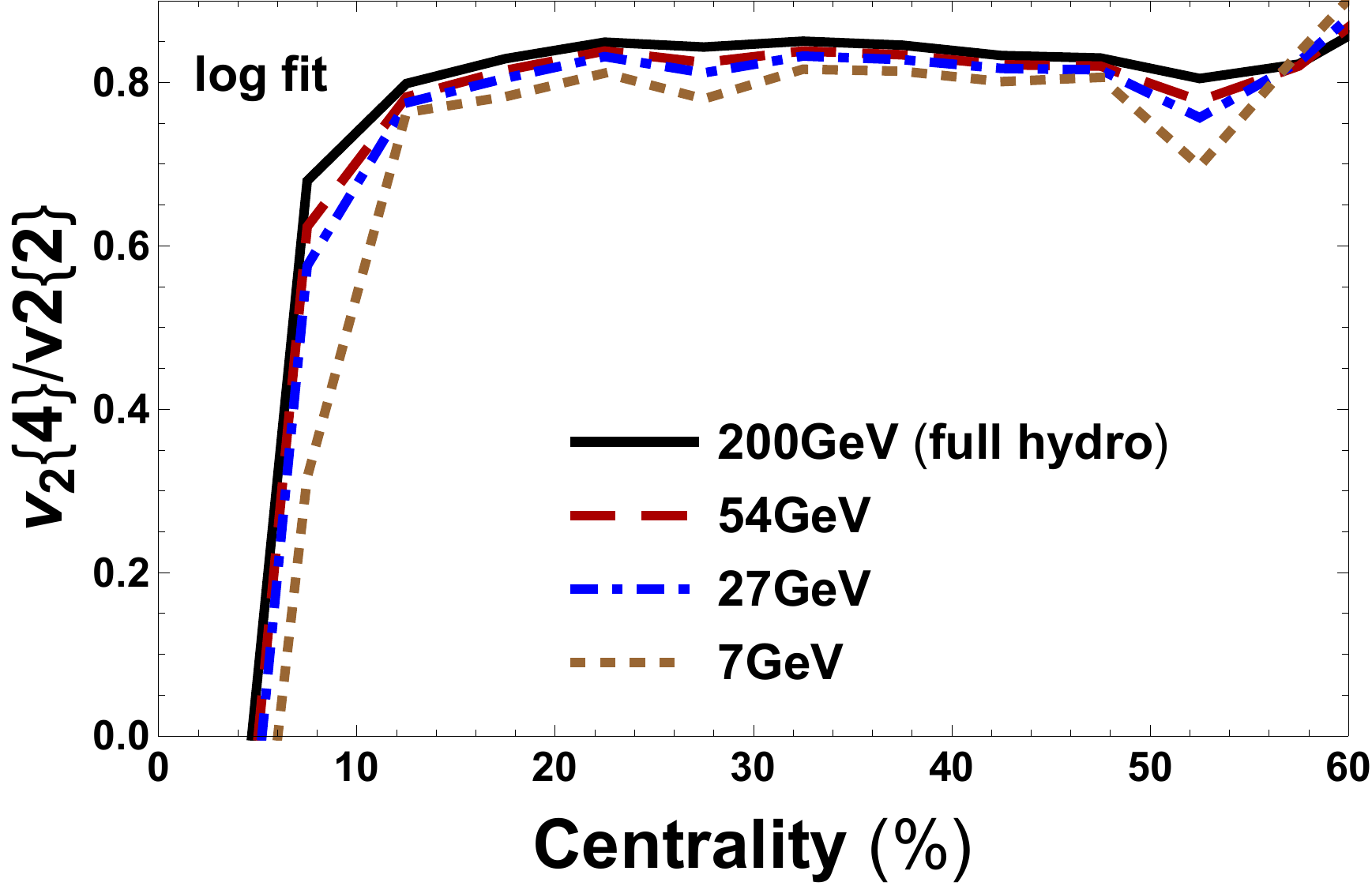} 
	\end{tabular}
	\caption{(Color online) Direct Trento+v-USPhydro hydrodynamic calculations for AuAu $200~\mathrm{GeV}$ (solid black) versus the predicted $v_2\{2\}$ (left) and $v_2\{4\}/v_2\{2\}$ (right) from extrapolating the linear+cubic response coefficients and residual to lower energies.  Here we assume an logarithmic extrapolation $a+b\log[\sqrt{s_{NN}}]$ designed to mimic the onset of significant sub-eikonal physics at lower energies. Predictions are made for $54~\mathrm{GeV}$ (red long dashed), $27~\mathrm{GeV}$ (blue dot dashed), and $7~\mathrm{GeV}$ (brown short dashed).}
	\label{fig:logpreds}
\end{figure}
%

However, given that the assumption of an equal number of baryons and anti-baryons at top RHIC and LHC energies must break down at lower energies and associated changes are anticipated in the equation of state and transport coefficients, it is perhaps more realistic to anticipate the onset of a much stronger energy dependence at the lower RHIC beam energies.  Accordingly, in Fig.\ \ref{fig:logpreds} we allow the response coefficients to follow the toy logarithmic extrapolation, which allows for a significant suppression of $v_2\{2\}$ with decreasing beam energy.  These results appear to be much more consistent with the preliminary results from STAR \cite{Magdy:2018itt}; that data may provide early evidence for a dramatic change in the behavior of the response coefficients at lower beam energies associated medium effects at finite baryon densities.  However, even with this significant change in the behavior at lower energies, we see in the right panel of Fig.\ \ref{fig:logpreds} that the effects largely cancel for the cumulant ratio $v_2\{4\}/v_2\{2\}$, leading to only a slight suppression at lower beam energies.  This conclusion could change, however, with the addition of significant new effects in the initial state such as a change in the effective nucleon width $\sigma$ (see Fig.~\ref{fig:sigmabeam}) or in the final state such as changing transport coefficients.

%
\subsection{Extracting the Response Coefficients from Data}
\label{subsec:extractdata}
%

Finally, we explore the possibility of extracting the response coefficients directly from comparisons between initial conditions and data, without simulating hydrodynamics at all.  When the initial state eccentricities and the final-state flow harmonics are both calculated in theory, the expressions given in Eqs.~\eqref{eqn:nonlinear} provide the optimized estimator parameters which minimize the residuals.  But for a given model of the initial state, cumulants of the eccentricities $\varepsilon_n \{2\}$ and $\varepsilon_n \{4\} / \varepsilon_n \{2\}$ can also be directly compared with the measured cumulants of final-state flow in order to study the response of the system.  Moreover, the independent information provided by the two-particle cumulant $v_2 \{2\}$ and the cumulant ratio $v_2 \{4\} / v_2 \{2\}$ can help to constrain the estimator parameters (such as $\kappa_{1,2}, \kappa_{2,2}$ for linear+cubic response) which relate the measured flow to the chosen initial state model. For simplicity's sake, we do not vary the residuals since these are unknown factors in the mapping and additionally, we do not consider the residuals for $v_2 \{4\} / v_2 \{2\}$ because they are two separate sources ($\langle \delta_2\rangle$ and $\Delta_{2,4}$) of uncertainty, which mostly cancel out when the ratio is taken.

%
\begin{figure}[h]
	\centering
	\begin{tabular}{c c}
		\includegraphics[width=0.5\linewidth]{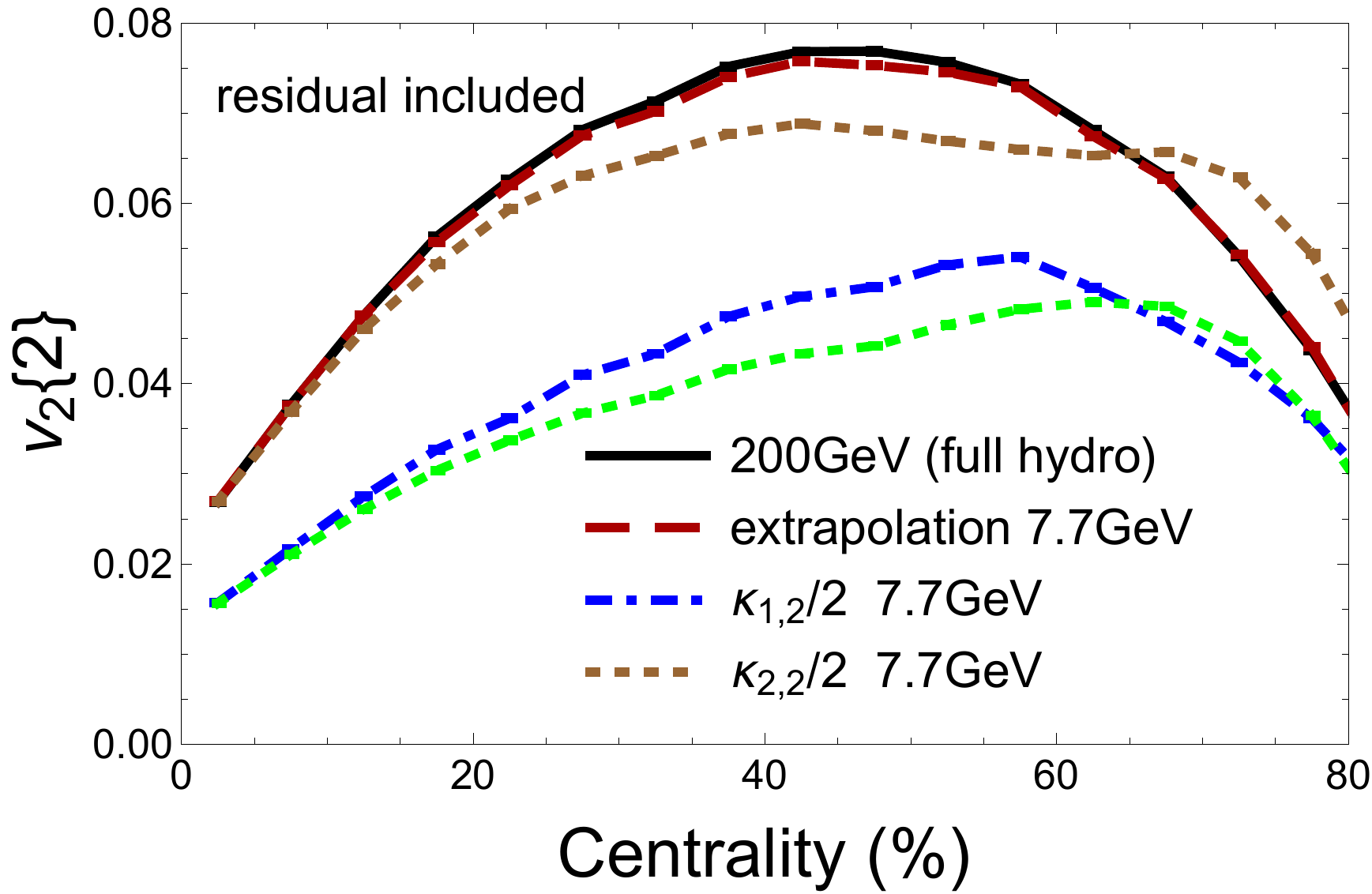} & \includegraphics[width=0.5\linewidth]{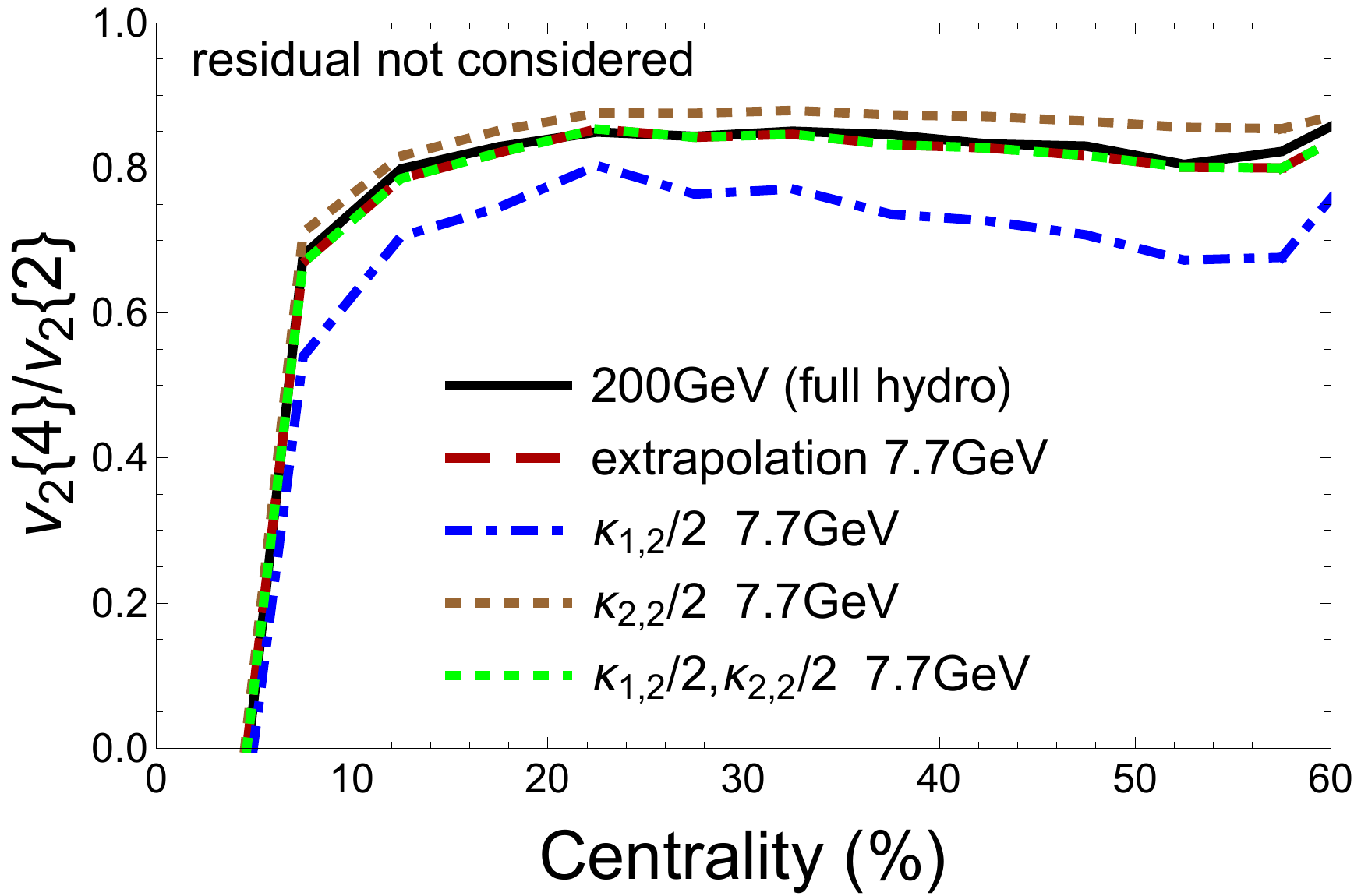} 
	\end{tabular}
	\caption{(Color online) Direct Trento+v-USPhydro hydrodynamic calculations (solid black) versus the predicted $v_n\{2\}$ from linear+cubic response (red long dashed) and the reconstructed $v_n\{2\}$ from linear+cubic response+residual (blue short dashed). Calculations at PbPb 5.02TeV.
	}
	\label{fig:half}
\end{figure}
%

Consider the impact of varying these parameters separately to predict AuAu $\sqrt{s_{NN}}=7.7$ GeV, as shown in Fig.~\ref{fig:half}.  We compare our linear extrapolation using the 3-energy fit (red long dashed line) with the prediction if the linear response coefficient was halved, $\kappa_{1,2}\rightarrow \kappa_{1,2}/2$ (blue dashed line).  Modifying the linear response coefficient in this way significantly decreases $v_2 \{2\}$, making it more in line with the STAR preliminary results \cite{Magdy:2018itt}.  However, at the same time, this change also significantly decreases the cumulant ratio $v_2 \{4\} / v_2 \{2\}$, which {\it{disagrees}} with the preliminary STAR data.  On the other hand, if we modify only the cubic response coefficient by a factor of $2$, $\kappa_{2,2}\rightarrow \kappa_{2,2}/2$, we see that there is actually a small {\it{increase}} in the cumulant ratio $v_2 \{4\} / v_2 \{2\}$.  This change may in fact be in agreement with the preliminary STAR data, but the error bars there are still large enough that it is difficult to say for sure.  However, now the opposite problem results: when halving the cubic response, there is only a mild decrease in $v_2 \{2\}$, which is likely still too large compared to the preliminary STAR results.  Finally, if both linear and cubic response coefficients are halved, then we find both a significant suppression of $v_2 \{2\}$ and a $v_2 \{4\} / v_2 \{2\}$ which is relatively unchanged.  This behavior appears to be similar in both observables to what is seen in the preliminary STAR results (and, interestingly, also to the predictions of our toy logarithmic extrapolation shown in Fig.\ \ref{fig:logpreds}).  

Taken together, these simple exercises suggest that it is feasible to constrain the linear and cubic response coefficients by directly comparing a model of the initial state to data.  Once the final STAR data is published and available to the public, these types of studies could be quite useful in determining the type of flow response to the initial state one expects at different beam energies.  Additionally, because these parameters encode the information about the final-state medium response to an initial state geometry, if they can be constrained directly from data then they can shed light on the new physical mechanisms which can be driving the change in the system.  Future studies varying the choices of medium properties such as the equation of state, transport coefficients, and hydrodynamic expansion time could compute their impact on the experimentally-constrained response coefficients to help extract what physics is driving the change in system response at lower beam energies.

%
\section{Conclusions}
\label{sec:concl}
%

In this paper we have studied the energy dependence of various parameterizations of the initial conditions of heavy ion collisions, finding almost no change in the eccentricities with energy.  This null result, however, is assuredly an artifact of the gluon-dominated physics of top RHIC and LHC energies being hard-coded into the various models through their underlying assumptions.  Deviations from this underlying physics associated with lower beam energies can change the picture of the initial state, for instance through additional changes in model parameters like the nucleon width or multiplicity fluctuations.  We found that changes in such secondary parameters with beam energy can have a mild effect on the initial-state eccentricities.  It is unclear how more dynamical approaches such as that of Ref.~\cite{Shen:2017bsr} would affect the initial eccentricities -- or even if a well-defined initial eccentricity could be constructed for such a scenario -- but we leave these considerations for future work.

We have also extracted the linear+cubic response coefficients across top RHIC and LHC energies using two different initial condition models: Trento $p=0$ which approximates the eccentricities produced by the IP-Glasma model, and MC-KLN.  Using these response coefficients, we extrapolated down to lower beam energies using either a linear or logarithmic fit to make baseline predictions from which to measure the expected deviations.  As expected, a linear fit to top beam energies showed almost no $\sqrt{s_{NN}}$ dependence for $v_2\{2\}$, in contrast to preliminary STAR results \cite{Magdy:2018itt}.  This scenario essentially assumes that the physics of top RHIC and LHC energies will continue unabated down to small $\sqrt{s_{NN}}$, which must surely be wrong at some finite energy.  

In contrast, since we expect finite baryon densities to lead to significant changes in the medium (and thus the medium response coefficients), a more severe extrapolation to low energies may be appropriate.  To this end, we used a toy logarithmic extrapolation of the response parameters down from top collider energies, leading to a more significant suppression in $v_2\{2\}$ while leaving the cumulant ratio $v_2 \{4\} / v_2 \{2\}$ nearly unchanged; this scenario appears to be more in line with the preliminary STAR results.  We also tested various approaches to the extraction these response coefficients, finding that the event plane angles can have significant effects on the quality of the predictions, so we emphasize that Eqs.\ \ref{eqn:nonlinear} should be used to extract the linear+cubic response coefficients instead of numerical techniques which take into account only the magnitudes. 

Comparing results from STAR and PHENIX, we find that there are hints of differences in $v_2\{4\}/v_2\{2\}$ if subevents are used to remove non-flow effects.  However, since the centrality bins used for these calculations are quite fine the error bars are too large to determine this with confidence.  Additionally, the error bar are likely enhanced because we cannot take into account correlated error in our error propagation.  Thus, we would encourage experimentalists to determine the error bars for the ratio $v_2\{4\}/v_2\{2\}$ (as was done for 5.02TeV in ATLAS \cite{Aad:2019fgl}) to determine the effect of subevents.

We also explored the ability to extract or constrain the response coefficients to a given initial state model by direct comparison to experimental data.  In the case of the STAR preliminary data \cite{Magdy:2018itt}, a suppression of both the linear and cubic response coefficients appear to be necessary in order to be in the right ballpark; if only one of these coefficients is suppressed then this would fail the constraints of simultaneously fitting both $v_2\{2\}$ and $v_2\{4\}/v_2\{2\}$.  When the final STAR data becomes public, we can refine this method to directly extract the response coefficients from the experimental data and set the stage for further modeling to understand the microscopic origin of the change in system response, including the equation of state, transport coefficients, and lifetime of hydrodynamics.  This approach provides an exciting opportunity to disentangle the changes in the underlying physics with beam energy and may even help to determine the influence a critical point could have on the measured flow harmonics.

\section*{Acknowledgments}
The authors thank Chun Shen and Bjoern Schenke for providing the IP-Glasma+MUSIC results shown in this paper.
J.N.H. acknowledges the support of the Alfred P. Sloan Foundation, support from the US-DOE Nuclear Science Grant No. DE-SC0019175, and the Office of Advanced Research Computing (OARC) at Rutgers, The State University of New Jersey for providing access to the Amarel cluster and associated research computing resources that have contributed to the results reported here.

\section*{References}


%

\end{document}